\documentclass[a4paper,12pt]{article}
\usepackage{amsmath,enumerate}
\usepackage{graphicx}

\newcommand{\lapprox}{%
\mathrel{%
\setbox0=\hbox{$<$}
\raise0.6ex\copy0\kern-\wd0
\lower0.65ex\hbox{$\sim$}
}}
\newcommand{\gapprox}{%
\mathrel{%
\setbox0=\hbox{$>$}
\raise0.6ex\copy0\kern-\wd0
\lower0.65ex\hbox{$\sim$}
}}

\begin{document}

\begin{center}

{\Large \bf Neutrino mixing by modifying the Yukawa coupling structure
of constrained sequential dominance}\\[20mm]

Joy Ganguly\footnote{ph18resch11009@iith.ac.in} and
Raghavendra Srikanth Hundi\footnote{rshundi@phy.iith.ac.in}\\
Department of Physics, Indian Institute of Technology Hyderabad,\\
Kandi - 502 285, India.\\[20mm]

\end{center}

\begin{abstract}

In the constrained sequential dominance (CSD), tri-bimaximal mixing (TBM)
pattern in the neutrino sector has been explained, by proposing a certain
Yukawa coupling structure for the right-handed neutrinos of the model.
However, from the current experimental data it is known that the values
of neutrino mixing angles are deviated from the TBM values. In order to
explain this neutrino mixing, we first propose a phenomenological model
where we consider Yukawa couplings which are modified from that of CSD.
Essentially, we add small complex parameters to
the Yukawa couplings of CSD. Using these modified Yukawa couplings, we
demonstrate that neutrino mixing angles can deviate from their TBM values.
We also construct a model, based on a flavor symmetry, in order to justify
the modified form of Yukawa couplings of our work.

\end{abstract}


\newpage
\section{Introduction}

From various experimental observations it is known that neutrinos have
very small mass \cite{rev}.
In a Type I seesaw mechanism, through the mediation of
heavy right-handed neutrinos, smallness of neutrino masses can be
understood \cite{sesaw,ts1}. To test this mechanism at
the LHC the mass of the right-handed neutrinos should be around 1 TeV. However,
with 1 TeV masses for right-handed neutrinos some tuning in the
Yukawa couplings may be required in order to fit the tiny masses
of neutrinos. Moreover, due to large number of seesaw parameters
this mechanism may not be predicted from the experimental data.
To alleviate the above
mentioned problems, models based on sequential dominance \cite{sd,dev-sd}
with two right-handed
neutrinos and one texture zero in the neutrino Yukawa matrix have been
proposed \cite{csd,csdn}. These models are named as CSD($n$),
which we describe them briefly below.

It is known that the neutrinos mix among them \cite{rev} and the
current oscillation data \cite{glo-fit} suggest that the neutrino mixing
angles are close to the TBM pattern \cite{tbm}. To explain these
mixing angles in the models of CSD($n$), the two
right-handed neutrinos are proposed to have certain particular Yukawa
couplings with the three lepton doublets. To be specific, the two right-handed
neutrinos, up to proportionality factors, are proposed to have the following
Yukawa couplings: $(0,1,1)$ and $(1,n,n-2)$.
Here, $n$ is a positive integer but can be taken to be
real as well. For the case of $n=1$, the model predicts that the three mixing
angles will take the following TBM values:
$\sin\theta_{12}=\frac{1}{\sqrt{3}}$, $\sin\theta_{23}=\frac{1}{\sqrt{2}}$,
$\sin\theta_{13}=0$. This case of $n=1$ is originally named as constrained
sequential dominance (CSD), which was viable a decade ago. But this
case has been ruled out when Daya Bay and RENO measured the $\theta_{13}$
and found it to be non-zero \cite{nu-ex}.
Among the other integer values for $n$, only the models with
$n=3,4$ are compatible with the current neutrino oscillation data \cite{csdn}.

In this work, we study on a possibility where we consider modifications
to model parameters of CSD and demonstrate that the neutrino observables
from the oscillation data can be explained. As explained above that CSD
is nothing but CSD($n=1$) and hence the Yukawa couplings in this model
are proportional to $(0,1,1)$ and $(1,1,-1)$. In the next section we will
describe that with this particular form for Yukawa couplings, the mixing
angles for neutrinos can be predicted to have the TBM values. Now, in order
to get deviations in neutrino mixing angles away from the TBM values,
we consider the Yukawa couplings of the two right-handed neutrinos
to be proportional to $(\epsilon_1,1+\epsilon_2,1+\epsilon_3)$
and $(1+\epsilon_4,1+\epsilon_5,-1+\epsilon_6)$. Here,
$\epsilon_i,i=1,\cdots,6$, are complex numbers. By proposing above mentioned
Yukawa couplings for neutrinos, we are considering here a phenomenological
model. Now, in this phenomenological model,
in the limit where all $\epsilon_i\to 0$,
our model should give the results of CSD. As a result of this,
we can expect that for small parametric values of $\epsilon_i$ we
should get deviations in neutrino mixing angles away from the TBM values.
The reason for considering all $\epsilon_i$ to be small is due to
the fact that the observed mixing angles are close to the TBM values.
After assuming that $\epsilon_i$ to be small, we study if we can consistently
fit the neutrino masses and mixing angles, whose values are obtained from
oscillation data.

Like in the model of CSD, in our model also only
two right-handed neutrinos are proposed.
As a result of this, in our model, one
neutrino would be massless and the other two can have non-zero masses.
Hence, in this model, we will show that only normal hierarchy is possible for
neutrino masses. We can fit the non-zero masses of our model to square
root of solar ($\sqrt{\Delta m^2_{\rm sol}}$) and atmospheric
($\sqrt{\Delta m^2_{\rm atm}}$) mass squared differences.
From the global fits to
neutrino oscillation data we can see that there is a hierarchy between
$\Delta m^2_{\rm sol}$ and $\Delta m^2_{\rm atm}$ \cite{glo-fit}. In fact,
from the results of ref.\cite{glo-fit}, one can notice that
$\frac{\Delta m^2_{\rm sol}}{\Delta m^2_{\rm atm}}\sim\sin^2\theta_{13}
\approx 2\times10^{-2}$. Because of this, we take
$\sqrt{\frac{\Delta m^2_{\rm sol}}{\Delta m^2_{\rm atm}}}$ and
$\sin\theta_{13}$ to be small, whose values can be around 0.15.

As mentioned above, in our work, we are modifying the neutrino Yukawa
couplings of CSD model by introducing small complex $\epsilon_i$ parameters.
To be
consistent with the oscillation data, we assume that the magnitude of
real and imaginary
parts of $\epsilon_i$ to be less than or of the order of
$\sqrt{\frac{\Delta m^2_{\rm sol}}{\Delta m^2_{\rm atm}}}\sim\sin\theta_{13}$.
After assuming this, we diagonalize the seesaw formula for active neutrinos
in our model, by following an approximation procedure, where we expand the
seesaw formula in power series of $\epsilon_i$. Recently, this kind of
diagonalization procedure has been used in a different neutrino mass
model \cite{rel}. In the context of this present work, the usage and
relevance of this diagonalization procedure have been described in
sections 3 and 4. Following this
diagonalization procedure, we derive expressions for neutrino masses and
mixing angles in terms of $\epsilon_i$. We show that by keeping terms
up to first order in $\epsilon_i$ of our analysis, we get $\sin\theta_{13}$ and
and $\sin\theta_{23}-\frac{1}{\sqrt{2}}$ to be non-zero but
$\sin\theta_{12}-\frac{1}{\sqrt{3}}$ is found to be undetermined.
In order to know if $\sin\theta_{12}-\frac{1}{\sqrt{3}}$ can be determined,
we compute expressions
in our analysis up to second order in $\epsilon_i$. Thereafter we demonstrate
that $\sin\theta_{12}-\frac{1}{\sqrt{3}}$ can also
be determined by $\epsilon_i$ parameters. Using the analytic expressions
for neutrino masses and mixing angles, in order to be compatible with current
neutrino oscillation data, we present numerical results and also
demonstrate that the assumptions made in our diagonalization procedure
are viable.

We study the above described work in a phenomenological model, where the
neutrino Yukawa couplings of this model are modified from that of CSD
model. One would like to know how such modified form for Yukawa couplings could
be possible in our model. In order to address this point, towards the
end of this paper, we construct a model, based on symmetry groups,
where we explain the
smallness of $\epsilon_i$ parameters and also
justify the structure of Yukawa couplings of our phenomenological model.
In order for this model to explain the structure of Yukawa couplings,
the scalar fields proposed in this model need to acquire vacuum expectation
values (vevs) with hierarchically different magnitudes. To explain the hierarchy
in the vev of these scalar fields, we analyze the scalar potential among these
fields and give one solution to this problem.

Deviations from TBM pattern has been studied in sequential dominance
models \cite{dev-sd}, where neutrino masses and mixing angles are computed in a
general framework of type I seesaw model and then these results are applied
to models which satisfy sequential dominance conditions. Here, our approach
to the problem is different from that of ref.\cite{dev-sd}. In this work,
we first modify the Yukawa coupling
structure of CSD and then study the deviations from TBM pattern. Moreover,
our analysis is also different from that of ref.\cite{dev-sd}.

The paper is organized as follows. In the next section we describe sequential
dominance and the CSD model. In section 3, we describe our phenomenological
model and also explain the approximation procedure for diagonalizing
the seesaw formula for neutrinos of this model. Using this approximation
procedure we demonstrate that the neutrino mixing angles in our model deviate
away from the TBM pattern. In the same section, we compute
expressions for neutrino masses and mixing angles up to first order
in our approximation scheme. Second order corrections to the above mentioned
neutrino observables have been computed in section 4. In section 5, we
give numerical results where we demonstrate that our analytic expressions
can fit the
current neutrino oscillation data. In section 6, we
construct a model in order to justify the structure of Yukawa
couplings of our phenomenological model. We conclude in the last section.
In Appendix A, we have
given detailed expressions related to the second order corrections to the
neutrino observables. In Appendix B, we analyze the scalar potential
of our model in order to explain the hierarchy in the vevs of the scalar fields.

\section{Sequential dominance and CSD}

The idea for CSD is motivated from sequential dominance, which is briefly
described below. Consider a minimal extension to the standard model,
where the additional fields are three singlet right-handed neutrinos. After
electroweak symmetry breaking, charged leptons and neutrinos acquire
mixing mass matrices. We can consider a basis in which both charged
leptons and right-handed neutrinos have been diagonalized. In this basis,
the mass matrix for right-handed neutrinos and the mixing mass matrix between
left- and right-handed neutrinos can be written, respectively, as
\begin{equation}
M_R = \left(\begin{array}{ccc}
M_{\rm atm} & 0 & 0\\ 0 & M_{\rm sol} & 0\\ 0 & 0 & M_{\rm dec}
\end{array}\right), \quad
m_D = \left(\begin{array}{ccc}
d & a & a^\prime \\ e & b & b^\prime \\ f & c & c^\prime
\end{array}\right)
\end{equation}
In the equation for $m_D$, elements such as $a,b,c$, etc can be viewed
as neutrino Yukawa coupling multiplied by vev of the
Higgs field. Assuming that the masses for right-handed neutrinos are much
larger than the elements of Dirac mass matrix, the seesaw formula
for active neutrinos would be
\begin{equation}
m_\nu = m_DM_R^{-1}m_D^T
\label{E:sesa}
\end{equation}
From the seesaw formula we get three masses for active neutrinos, which
may be denoted by $m_1$, $m_2$ and $m_3$. The objective of sequential
dominance is to achieve $m_1\ll m_2\ll m_3$, and thereby the model
can predict normal mass hierarchy for neutrinos. In order to achieve
this objective of sequential dominance, following assumptions
on the masses of right-handed neutrinos and the elements of the Dirac
mass matrix have been made \cite{sd,dev-sd}
\begin{equation}
M_{\rm atm}\ll M_{\rm sol}\ll M_{\rm dec},\quad
\frac{|e^2|,|f^2|,|ef|}{M_{\rm atm}}\gg\frac{xy}{M_{\rm sol}}\gg
\frac{x^\prime y^\prime}{M_{\rm dec}}
\end{equation}
Here, $x,y\in a,b,c$ and $x^\prime,y^\prime\in a^\prime,b^\prime,c^\prime$.

With the above mentioned assumptions of sequential dominance, leading
order expressions for neutrino masses and mixing angles have been
computed in ref.\cite{led-ex}. Using these expressions,
following set of conditions on the model parameters have been proposed,
in order to obtain the TBM pattern for neutrino mixing angles \cite{csd}.
\begin{equation}
|a| = |b| =|c|,\quad |d| = 0, \quad |e| = |f|,\quad
\phi_b^\prime = 0,\quad \phi_c^\prime = \pi
\label{E:cond}
\end{equation}
Here, $\phi_b^\prime$ and $\phi_c^\prime$ denote sum of a combination
of phases of the
elements in the Dirac mass matrix \cite{csd}. From the above mentioned
conditions we can notice that the elements in the third column of
$m_D$ and $M_R$ play no part in determining the TBM pattern for
neutrino mixing angles. In fact, from the leading order expressions
for neutrino masses and mixing angles given in ref.\cite{led-ex}, we
can see that the third column elements of $m_D$ and $M_R$ determine only
the lightest neutrino mass $m_1$. One can notice that $m_1$ is proportional
to $\frac{1}{M_{\rm dec}}$. Now, in the limit where the value of $M_{\rm dec}$
tends to very large, we get $m_1\to 0$. In this limiting process, the third
right-handed neutrino, whose mass is $M_{\rm dec}$, decouples from our
theory.
Since the current experimental data can be satisfied with $m_1=0$,
in order to reduce
the number of degrees of freedom in this model, we can decouple away the
third column
elements of $m_D$ and $M_R$. Essentially, in this process of decoupling,
the number of right-handed neutrinos reduce from three to two in the above
described model.

After performing the above mentioned decoupling, in the resultant model,
to satisfy the conditions of
Eq. (\ref{E:cond}), the Dirac and right-handed neutrino mass matrices
can be taken, respectively, as \cite{csd}
\begin{equation}
m_D = \left(\begin{array}{cc}
0 & a \\ e & a \\ e & -a
\end{array}\right),\quad
M_R = \left(\begin{array}{cc}
M_{\rm atm} & 0 \\ 0 & M_{\rm sol}
\end{array}\right)
\end{equation}
By plugging the above mentioned $m_D$ and $M_R$ in the seesaw formula
of Eq. (\ref{E:sesa}), we can check that the $m_\nu$ can be diagonalized
as
\begin{equation}
U_{\rm TBM}^Tm_\nu U_{\rm TBM} = \left(\begin{array}{ccc}
0 & 0 & 0\\ 0 & \frac{3a^2}{M_{\rm sol}} & 0 \\ 0 & 0 &
\frac{2e^2}{M_{\rm atm}}
\end{array}\right),\quad
U_{\rm TBM} = \left(\begin{array}{ccc}
\sqrt{\frac{2}{3}} & \frac{1}{\sqrt{3}} & 0 \\
-\frac{1}{\sqrt{6}} & \frac{1}{\sqrt{3}} & \frac{1}{\sqrt{2}} \\
\frac{1}{\sqrt{6}} & -\frac{1}{\sqrt{3}} & \frac{1}{\sqrt{2}}
\end{array}\right)
\label{eq:tbm}
\end{equation}
From the unitary matrix $U_{\rm TBM}$, one can extract the three
neutrino mixing angles and we see that they will have the TBM values.

We have demonstrated above that in a model with two right-handed neutrinos,
which is motivated by sequential dominance,
TBM pattern for neutrino mixing is
possible. This has been named as CSD \cite{csd}. One can notice that in
this process of obtaining TBM pattern, the columns of Dirac mass matrix need
to be aligned in some particular directions. This problem of alignment
has been addressed in a supersymmetric model
which has some flavor symmetries and flavon fields \cite{csd}.

\section{Our model and deviations from TBM pattern}

In the previous section we have described on how CSD can predict TBM pattern for
neutrino mixing angles. Since this pattern is currently ruled out,
we need to modify the model of CSD. To achieve this, we initially consider
a phenomenological model where the field content is same as that of CSD.
But the difference between our model and the CSD is that we propose a
modified structure for Dirac mass matrix, which is given below.
\begin{equation}
m_D^\prime = m_D + \Delta m_D,\quad
m_D = \left(\begin{array}{cc}
0 & a \\ e & a \\ e & -a
\end{array}\right),\quad
\Delta m_D = \left(\begin{array}{cc} e\epsilon_1 & a\epsilon_4 \\
e\epsilon_2 & a\epsilon_5 \\ e\epsilon_3 & a\epsilon_6 \end{array}\right)
\label{eq:mdd}
\end{equation}
Here, $\epsilon_i,i=1,\cdots,6$, are complex parameters. At this stage we
are suggesting the above form for Dirac mass matrix, purely from
phenomenological point of view. We justify this form of matrix by
constructing a model for this in section 6.
Regarding the Dirac mass matrix, we have explained in the previous
section that the elements of this matrix should be viewed as a product
of neutrino Yukawa couplings and vev of the Higgs field. As a result of
this, the above Dirac mass matrix corresponds to the fact that the Yukawa
couplings of the two right-handed neutrinos are proportional to
$(\epsilon_1,1+\epsilon_2,1+\epsilon_3)$ and
$(1+\epsilon_4,1+\epsilon_5,-1+\epsilon_6)$. As we have argued in section 1,
with this form for Yukawa couplings we should expect to get deviations for
neutrino mixing angles away from the TBM values.

As explained above that in our model, the form for Dirac mass matrix is
given by $m_D^\prime$ and hence the seesaw formula for active neutrinos
is
\begin{equation}
m^s_\nu = m_D^\prime M_R^{-1}(m^\prime_D)^T
\label{eq:ours}
\end{equation}
Since we are in a basis where
charged leptons are diagonalized, this seesaw formula should be diagonalized
by Pontecorvo-Maki-Nakagawa-Sakata (PMNS) matrix. The PMNS matrix can
be parametrized by the neutrino mixing angles and the CP violating Dirac
phase $\delta_{\rm CP}$. We follow the PDG convention for this parametrization
\cite{pdg}, which is given below.
\begin{equation}
U_{\rm PMNS} = \left(\begin{array}{ccc}
c_{12}c_{13} & s_{12}c_{13} & s_{13}e^{-i\delta_{\rm CP}} \\
-s_{12}c_{23}-c_{12}s_{23}s_{13}e^{i\delta_{\rm CP}} &
c_{12}c_{23}-s_{12}s_{23}s_{13}e^{i\delta_{\rm CP}} & s_{23}c_{13} \\
s_{12}s_{23}-c_{12}c_{23}s_{13}e^{i\delta_{\rm CP}} &
-c_{12}s_{23}-s_{12}c_{23}s_{13}e^{i\delta_{\rm CP}} & c_{23}c_{13}
\end{array}\right)
\label{eq:pmns2}
\end{equation}
Here, $c_{ij}=\cos\theta_{ij}$ and $s_{ij}=\sin\theta_{ij}$. As explained
above that in our model, with the form for $m_D^\prime$ of Eq. (\ref{eq:mdd}),
we should get deviations in the neutrino mixing angles away from the
TBM values. As a result of this, we should expect $s_{13}$,
$s_{12}-1/\sqrt{3}$ and $s_{23}-1/\sqrt{2}$ to become non-zero. In order
to simplify our calculations, we parametrize $s_{12}$ and $s_{23}$ as
\begin{equation}
s_{12}=\frac{1}{\sqrt{3}}(1+r),\quad s_{23}=\frac{1}{\sqrt{2}}(1+s)
\end{equation}
The parametrization we have considered for neutrino mixing angles is
similar to that proposed in refs.\cite{para}. For a different parametrization
of these neutrino mixing angles, see ref.\cite{d-para}.
We have known the 3$\sigma$ ranges for the square of the sine of
the neutrino mixing angles, which are obtained from the global fits to
oscillation data \cite{glo-fit}. From these 3$\sigma$ ranges, we can find
the corresponding ranges for $r$ and $s$, which are found,
respectively, as:
$(-8.8\times 10^{-2},2.5\times 10^{-2})$ and $(-8.2\times 10^{-2},0.13)$.
The corresponding allowed range for $s_{13}$ is found to be narrow, whose
values are around 0.15. From the above mentioned ranges, we can notice that
the values for $r$ and $s$ are less than or of the order of $s_{13}$.
As explained before that $r$, $s$ and $s_{13}$ will become non-zero
in our model, if we allow non-zero values for $\epsilon_i$ parameters in
$m_D^\prime$. As a result of this, to be consistent with our analysis, we
assume that the real and imaginary parts of $\epsilon_i$ to
be less than or of the order of $s_{13}$.

As described previously, seesaw formula for active neutrinos in our model
is given by Eq. (\ref{eq:ours}) and this matrix should be
diagonalized by $U_{\rm PMNS}$. The relation for this diagonalization
can be written as
\begin{equation}
m_\nu^d\equiv U_{\rm PMNS}^T m^s_\nu U_{\rm PMNS}
= {\rm diag}(m_1,m_2,m_3)
\label{eq:md}
\end{equation}
Here, the matrices $m_\nu^s$ and $U_{\rm PMNS}$ depend on variables
$\epsilon_i$, $r$, $s$ and $s_{13}$, which are small. As a result of this,
we can expand $m_\nu^s$ and $U_{\rm PMNS}$ as power series in terms of these
small variables. First we expand $m_\nu^s$ and $U_{\rm PMNS}$ up to first
order in $\epsilon_i$, $r$, $s$ and $s_{13}$.
After doing that one can see that $m_\nu^d$ need not
be in diagonal form. But, since we expect this to be of diagonal form, we
demand that the off-diagonal elements of $m_\nu^d$ to be zero. Thereby we
get three relations among $\epsilon_i$, $r$, $s$ and $s_{13}$. Solving
these relations, we can determine $\epsilon_i$ in terms of $r$, $s$ and
$s_{13}$. Now, from the diagonal elements of $m_\nu^d$ we get
expressions for the three neutrino masses in terms of model parameters. 
We follow the above described methodology for diagonalizing the seesaw
formula of our model. However, while doing so, one needs to take care of
the small numbers that may arise due to hierarchy in neutrino masses.
Discussion related to this is explained below.

In the limit where $\epsilon_i$, $r$, $s$ and $s_{13}$ tend to zero, from
Eq. (\ref{eq:md}) we get the leading order expressions for neutrino masses,
which are given below.
\begin{equation}
m_1=0,\quad m_2=\frac{3a^2}{M_{\rm sol}},\quad m_3=\frac{2e^2}{M_{\rm atm}}
\label{eq:led}
\end{equation}
The above result agree with that of CSD which is given in section 2. Here,
up to the leading order, the lightest neutrino mass $m_1$ is zero. However, we
will show later that even at sub-leading orders, $m_1$ is still zero. This
result is due to the consequence of the fact that in our model we have
proposed only two right-handed neutrinos. As a result of this, neutrino
masses in our model can only have normal mass hierarchy. Due to this, we
can fit the expressions for $m_2$ and $m_3$ to square root of solar
($\sqrt{\Delta m^2_{\rm sol}}$) and atmospheric ($\sqrt{\Delta m^2_{\rm atm}}$)
mass squared differences, respectively.
Although the expressions in Eq. (\ref{eq:led}) are valid at
leading order, at sub-leading orders, expressions for $m_2$ and $m_3$
get corrections which are proportional to $\epsilon_i$, $r$, $s$ and $s_{13}$.
Since $\epsilon_i$, $r$, $s$ and $s_{13}$ are small values, when we fit the
expressions for $m_2$ and $m_3$ to
$\sqrt{\Delta m^2_{\rm sol}}$ and $\sqrt{\Delta m^2_{\rm atm}}$ respectively,
we except to have the following order of estimations.
\begin{equation}
\frac{a^2}{M_{\rm sol}}\sim\sqrt{\Delta m^2_{\rm sol}},\quad
\frac{e^2}{M_{\rm atm}}\sim\sqrt{\Delta m^2_{\rm atm}}
\label{eq:asum}
\end{equation}
We use the above mentioned order of estimations in the diagonalization
process of the seesaw formula of our model. Regarding this, a point to be
noticed here is
that, from the global fits to neutrino oscillation data \cite{glo-fit}, a
hierarchy is found between $\Delta m^2_{\rm sol}$ and $\Delta m^2_{\rm atm}$.
In fact, from the results of ref.\cite{glo-fit}, one can notice that
$\sqrt{\frac{\Delta m^2_{\rm sol}}{\Delta m^2_{\rm atm}}}\sim s_{13}$. This
would imply that, in our model, $m_2/m_3\sim s_{13}$. One needs to incorporate
the above mentioned order of estimation in the diagonalization process of
the seesaw formula of our model. In order to incorporate this, we reexpress
Eq. (\ref{eq:md}) as
\begin{equation}
\frac{1}{\sqrt{\Delta m^2_{\rm atm}}}m_\nu^d\equiv
\frac{1}{\sqrt{\Delta m^2_{\rm atm}}}U_{\rm PMNS}^T m^s_\nu U_{\rm PMNS}
= {\rm diag}(\frac{m_1}{\sqrt{\Delta m^2_{\rm atm}}},\frac{m_2}
{\sqrt{\Delta m^2_{\rm atm}}},\frac{m_3}{\sqrt{\Delta m^2_{\rm atm}}})
\label{eq:mdp}
\end{equation}
Now, with the assumptions of Eq. (\ref{eq:asum}), one can see that
$\frac{1}{\sqrt{\Delta m^2_{\rm atm}}}m_\nu^d$ can be expanded in power
series of $\epsilon_i$, $r$,
$s$, $s_{13}$ and $\sqrt{\frac{\Delta m^2_{\rm sol}}{\Delta m^2_{\rm atm}}}$.
We explain below about this series expansion and also the results obtained
from such expansion.

Up to first order in $\epsilon_i$, $m_\nu^s$ can be
expanded as
\begin{eqnarray}
m_\nu^s &=& m_{\nu(0)}^s + m_{\nu(1)}^{s},
\label{eq:ms1} \\
m_{\nu(0)}^{s} &=& m_DM_R^{-1}m_D^T,
\quad m_{\nu(1)}^{s} = m_DM_R^{-1}(\Delta m_D)^T + \Delta m_DM_R^{-1}m_D^T
\label{eq:mn01}
\end{eqnarray}
Similarly, up to first order in $r$, $s$ and $s_{13}$, the expansion for
$U_{\rm PMNS}$ is
\begin{eqnarray}
U_{\rm PMNS} &=& U_{\rm TBM} + \Delta U,
\label{eq:upmns1} \\
\Delta U &=& \left(
\begin{array}{ccc}
-\frac{r}{\sqrt{6}} & \frac{r}{\sqrt{3}} & e^{-i\delta_{\rm CP}}s_{13} \\
\frac{-r+s}{\sqrt{6}}-\frac{e^{i\delta_{\rm CP}}s_{13}}{\sqrt{3}} &
-\frac{r+2s+\sqrt{2}e^{i\delta_{\rm CP}}s_{13}}{2\sqrt{3}} &
\frac{s}{\sqrt{2}} \\
\frac{r+s}{\sqrt{6}}-\frac{e^{i\delta_{\rm CP}}s_{13}}{\sqrt{3}} &
\frac{r-2s-\sqrt{2}e^{i\delta_{\rm CP}}s_{13}}{2\sqrt{3}} &
-\frac{s}{\sqrt{2}}
\end{array}
\right)
\label{eq:du1}
\end{eqnarray}
Here, the form of $U_{\rm TBM}$ can be seen in Eq. (\ref{eq:tbm}).
After substituting Eqs. (\ref{eq:ms1}) $\&$ (\ref{eq:upmns1}) in Eq.
(\ref{eq:mdp}) and with the assumptions of Eq. (\ref{eq:asum}),
we can compute $\frac{1}{\sqrt{\Delta m^2_{\rm atm}}}m_\nu^d$
up to first order in $\epsilon_i$, $r$, $s$, $s_{13}$ and
$\sqrt{\frac{\Delta m^2_{\rm sol}}{\Delta m^2_{\rm atm}}}$. Terms up to
first order in $\frac{1}{\sqrt{\Delta m^2_{\rm atm}}}m_\nu^d$ are given below.
\begin{eqnarray}
&&\frac{1}{\sqrt{\Delta m^2_{\rm atm}}}m_\nu^d=
\frac{1}{\sqrt{\Delta m^2_{\rm atm}}}\left(m_{\nu(0)}^d+m_{\nu(1)}^d\right),
\nonumber \\
&&m_{\nu(0)}^d=\left(\begin{array}{ccc}
0 & 0 & 0\\ 0 & \frac{3a^2}{M_{\rm sol}} & 0 \\ 0 & 0 &
\frac{2e^2}{M_{\rm atm}}
\end{array}\right),\quad
m_{\nu(1)}^d=
\left(
\begin{array}{ccc}
x^\prime_{11} & x^\prime_{12} & x^\prime_{13} \\
x^\prime_{12} & x^\prime_{22} & x^\prime_{23} \\
x^\prime_{13} & x^\prime_{23} & x^\prime_{33}
\end{array}\right),
\nonumber \\
&& x^\prime_{11}=0,\quad x^\prime_{12}=0,\quad
x^\prime_{13}=\frac{e^2}{\sqrt{6}M_{\rm atm}}
[\sqrt{2}(2\epsilon_1-\epsilon_2+\epsilon_3+2s)-4e^{i\delta_{\rm CP}}s_{13}],
\quad x^\prime_{22}=0,
\nonumber \\
&& x^\prime_{23}=\frac{e^2}{\sqrt{3}M_{\rm atm}}
[\sqrt{2}(\epsilon_1+\epsilon_2-\epsilon_3-2s)
-2e^{i\delta_{\rm CP}}s_{13}],\quad
x^\prime_{33}=\frac{2e^2}{M_{\rm atm}}.
(\epsilon_2+\epsilon_3)
\end{eqnarray}
Now, equating the diagonal elements on both sides of Eq. (\ref{eq:mdp}),
we
get the expressions for the three neutrino masses, which are given below
\begin{equation}
m_1=0,\quad m_2=\frac{3a^2}{M_{\rm sol}},\quad m_3=\frac{2e^2}{M_{\rm atm}}+
\frac{2e^2(\epsilon_2+\epsilon_3)}{M_{\rm atm}}
\end{equation}
From the above equations we can see that only $m_3$ get correction at
the first order level.
Now, from the off-diagonal elements of Eq. (\ref{eq:mdp}), we get
the following expressions.
\begin{equation}
\epsilon_1 = \sqrt{2}e^{i\delta_{\rm CP}}s_{13},\quad \epsilon_2
-\epsilon_3=2s
\label{eq:res}
\end{equation}
From the above two equations we can see that, in our model, $\sin\theta_{13}$
will be non-zero if we take $\epsilon_1\neq 0$. Similarly, $\sin\theta_{23}$
will deviate from its TBM value if we take either $\epsilon_2$ or
$\epsilon_3$ to be non-zero. However, the deviation of $\sin\theta_{12}$
from its TBM value, which is quantified in terms of $r$, is undetermined
at the first order level corrections to the diagonalization of our seesaw
formula.
As a result of this, the parameters $\epsilon_4$, $\epsilon_5$ and $\epsilon_6$
are undetermined at this level. We will show in the next section that these
parameters can be determined in terms of neutrino mixing angles
by considering second order level
corrections to the diagonalization of our seesaw formula.

Results obtained in Eq. (\ref{eq:res}) are consistent with that in Partially
CSD (PCSD) \cite{pcsd}. In the model of PCSD, the structure of neutrino
Yukawa couplings is similar to that in our model. The Yukawa couplings in
PCSD can be obtained from that of our model by taking $\epsilon_1\neq 0$
and all other $\epsilon_i$ to be zero. With this Yukawa coupling structure,
in the
PCSD model, it is shown that $\sin\theta_{13}\neq 0$ after assuming TBM values
for $\sin\theta_{12}$ and $\sin\theta_{23}$. These results are obtained
in PCSD model up to a leading order in $m_2/m_3$. Results in previous paragraph
are also obtained up to to this order. Although we have argued that the
value of $r$ is undetermined up to this order, with out loss of generality,
in the beginning of the calculations,
we can assume TBM value for $\sin\theta_{12}$ and choose zero values for
$\epsilon_4$, $\epsilon_5$ and $\epsilon_6$.
In that case, we would still get the
results of Eq. (\ref{eq:res}). Now if we choose zero values for $\epsilon_2$
and $\epsilon_3$, that would imply TBM value for $\sin\theta_{23}$. Hence,
results obtained in the previous paragraph are consistent with that of
PCSD model. Moreover, it is to be noticed that the structure of our model
and the results obtained in this work generalizes that of PCSD model.
We have described above about the relevance of the relations in
Eq. (\ref{eq:res}).
It should be noted that, in our framework, these relations cannot be
obtained without making the assumptions of Eq. (\ref{eq:asum}). In the next
section we further stress on the usage of these assumptions and on the
consistency of the results obtained with our diagonalization
procedure.

\section{Second order corrections}

In the previous section, after considering first order
corrections to the diagonalization of the seesaw formula for neutrinos,
it is found that the deviation of $\sin\theta_{12}$ from its TBM
value is found to be undetermined. To know if this deviation can be determined
in terms of model parameters, we study here the second order corrections to
the diagonalization of the seesaw formula for neutrino masses.
In order to do this we need to
expand terms in $\frac{1}{\sqrt{\Delta m_{\rm atm}^2}}m_\nu^d$ of
Eq. (\ref{eq:mdp}) up to second order in $\epsilon_i$, $r$,
$s$, $s_{13}$ and $\sqrt{\frac{\Delta m^2_{\rm sol}}{\Delta m^2_{\rm atm}}}$.
Details related to this expansion and the analysis from that is explained
below.

Expansion for $m_\nu^s$ and $U_{\rm PMNS}$, up to second order in $\epsilon_i$, $r$, $s$ and $s_{13}$ are given below
\begin{eqnarray}
m_\nu^s &=& m_{\nu(0)}^s+m_{\nu(1)}^s+m_{\nu(2)}^s,\quad
m_{\nu(2)}^s = \Delta m_DM_R^{-1}(\Delta m_D)^T,
\\
U_{\rm PMNS} &=& U_{\rm TBM}+\Delta U+\Delta^2U,
\\
\Delta^2U &=& \left(\begin{array}{ccc}
-\frac{3r^2+4s_{13}^2}{4\sqrt{6}} & -\frac{s_{13}^2}{2\sqrt{3}} & 0 \\
\frac{\sqrt{2}(rs+s^2)+(r-2s)s_{13}e^{i\delta_{\rm CP}}}{2\sqrt{3}} &
\frac{-3r^2+4rs-8s^2-4\sqrt{2}(r+s)s_{13}e^{i\delta_{\rm CP}}}{8\sqrt{3}} &
-\frac{s_{13}^2}{2\sqrt{2}} \\
\frac{\sqrt{2}rs+(r+2s)s_{13}e^{i\delta_{\rm CP}}}{2\sqrt{3}} &
\frac{3r^2+4rs-4\sqrt{2}(r-s)s_{13}e^{i\delta_{\rm CP}}}{8\sqrt{3}} &
-\frac{2s^2+s_{13}^2}{2\sqrt{2}}
\end{array}\right)
\end{eqnarray}
Here, the expressions for $m_{\nu(0)}^s,m_{\nu(1)}^s$ and $\Delta U$
can be found in Eqs. (\ref{eq:mn01}) $\&$ (\ref{eq:du1}), while
$U_{\rm TBM}$ can be seen in Eq. (\ref{eq:tbm}).
After substituting the above described expansions for $m_\nu^s$ and
$U_{\rm PMNS}$ in Eq. (\ref{eq:mdp}), and also after using Eq. (\ref{eq:asum}),
$\frac{1}{\sqrt{\Delta m_{\rm atm}^2}}m_\nu^d$ can be computed up to
second order in $\epsilon_i$, $r$,
$s$, $s_{13}$ and $\sqrt{\frac{\Delta m^2_{\rm sol}}{\Delta m^2_{\rm atm}}}$.
The full expressions for second order terms in
$\frac{1}{\sqrt{\Delta m_{\rm atm}^2}}m_\nu^d$ are given in Appendix A.
Now, after using the results of Eq. (\ref{eq:res}) in Eq. (\ref{eq:2ndrel}),
the second order terms in $\frac{1}{\sqrt{\Delta m_{\rm atm}^2}}m_\nu^d$
will be simplified. These are given below.
\begin{eqnarray}
&&\frac{1}{\sqrt{\Delta m_{\rm atm}^2}}m_{\nu(2)}^d=
\frac{1}{\sqrt{\Delta m_{\rm atm}^2}}\left(
\begin{array}{ccc}
x^{\prime\prime}_{11} & x^{\prime\prime}_{12} & x^{\prime\prime}_{13} \\
x^{\prime\prime}_{12} & x^{\prime\prime}_{22} & x^{\prime\prime}_{23} \\
x^{\prime\prime}_{13} & x^{\prime\prime}_{23} & x^{\prime\prime}_{33}
\end{array}\right),
\nonumber \\
x^{\prime\prime}_{11}&=&0,\quad x^{\prime\prime}_{12}=\frac{a^2}
{\sqrt{2}M_{\rm sol}}(2\epsilon_4-\epsilon_5+\epsilon_6-3r),
\nonumber \\
x^{\prime\prime}_{13}&=&\frac{e^2}{\sqrt{3}M_{\rm atm}}[s(3s-2\sqrt{2}
e^{i\delta_{\rm CP}}s_{13})+2\epsilon_3(s-\sqrt{2}e^{i\delta_{\rm CP}}s_{13})],
\nonumber \\
x^{\prime\prime}_{22}&=&\frac{2a^2}{M_{\rm sol}}(\epsilon_4+\epsilon_5
-\epsilon_6),
\nonumber \\
x^{\prime\prime}_{23}&=&\frac{\sqrt{3}a^2}{2M_{\rm sol}}[\sqrt{2}
(\epsilon_5+\epsilon_6+2s)+2e^{-i\delta_{\rm CP}}s_{13}]
\nonumber \\
&&-\frac{e^2}{\sqrt{3}M_{\rm atm}}[2\epsilon_3(\sqrt{2}s+
e^{i\delta_{\rm CP}}s_{13})+s(3\sqrt{2}s+2e^{i\delta_{\rm CP}}s_{13})],
\nonumber \\
x^{\prime\prime}_{33}&=&\frac{2e^2}{M_{\rm atm}}(\epsilon_3^2+2
\epsilon_3s+2s^2+s_{13}^2)
\end{eqnarray}
Now, after equating the diagonal elements on both sides of Eq. (\ref{eq:mdp}),
we get corrections up to second order to neutrino masses, which are given
below.
\begin{eqnarray}
&& m_1=0,\quad m_2 = \frac{3a^2}{M_{\rm sol}}+\frac{2a^2}{M_{\rm sol}}
(\epsilon_4+\epsilon_5-\epsilon_6),
\nonumber \\
&& m_3 = \frac{2e^2}{M_{\rm atm}}+\frac{4e^2}{M_{\rm atm}}(\epsilon_3+s)
+\frac{2e^2}{M_{\rm atm}}(s_{13}^2+\epsilon_3^2+2\epsilon_3s+2s^2)
\label{eq:2om}
\end{eqnarray}
After demanding that the off-diagonal elements of
$\frac{1}{\sqrt{\Delta m_{\rm atm}^2}}m_\nu^d$ should be zero, we get the
following three relations.
\begin{eqnarray}
&& 2\epsilon_4-\epsilon_5+\epsilon_6 = 3r,
\label{eq:r}
\\
&& s(3s-2\sqrt{2}e^{i\delta_{\rm CP}}s_{13})+2\epsilon_3
(s-\sqrt{2}e^{i\delta_{\rm CP}}s_{13}) = 0,
\label{eq:e3}
\\
&& \sqrt{\frac{\Delta m^2_{\rm sol}}{\Delta m^2_{\rm atm}}}e^{i\phi}
[\sqrt{2}(\epsilon_5+\epsilon_6+2s)+2e^{-i\delta_{\rm CP}}s_{13}]
\nonumber \\
&&-[2\epsilon_3(\sqrt{2}s+e^{i\delta_{\rm CP}}s_{13})+s(3\sqrt{2}s+2
e^{i\delta_{\rm CP}}s_{13})] = 0.
\label{eq:3rd}
\end{eqnarray}
While obtaining Eq. (\ref{eq:3rd}), we have used the expressions for
$m_2$ and $m_3$ of Eq. (\ref{eq:2om}).
Here, $\phi$ is the Majorana phase difference in the neutrino masses
$m_2$ and $m_3$.

From the expressions for neutrinos masses which are given in Eq. (\ref{eq:2om}),
we can see that the lightest neutrino mass is $m_1=0$. As already explained
before, this result follows from the fact that there exists only two
right-handed neutrinos in our model. But technically, this result will follow
after using the relations of Eq. (\ref{eq:res}) in Eq. (\ref{eq:2ndrel}).
Since the relations in Eq. (\ref{eq:res}) are obtained after making the 
assumptions in Eq. (\ref{eq:asum}), we can notice here on
the consistency of the obtained results with our diagonalization
procedure,
which is described in the previous section.
It is stated above that the lightest neutrino mass is zero in our
model, and hence, only normal mass hierarchy is possible for neutrino masses.
As a result of this the expressions for $m_2$ and $m_3$ of Eq. (\ref{eq:2om})
can be fitted to $\sqrt{\Delta m_{\rm sol}^2}$
and $\sqrt{\Delta m_{\rm atm}^2}$ respectively. While doing this fitting,
we can notice that terms involving $\epsilon_i$, $s_{13}$ and $s$ give
small corrections. Hence, we can see that $\frac{a^2}{M_{\rm sol}}$ and
$\frac{e^2}{M_{\rm atm}}$ can be of the order of $\sqrt{\Delta m_{\rm sol}^2}$
and $\sqrt{\Delta m_{\rm atm}^2}$ respectively. This result agrees with
the assumption we have made in Eq. (\ref{eq:asum}). Another point to be
noticed here is that both the expressions for $m_2$ and $m_3$ depend
on the complex $\epsilon_i$ parameters. As a result of this, both $m_2$
and $m_3$ can be complex. But since neutrino masses should be real, the
complex phases in $m_2$ and $m_3$ can be absorbed into Majorana
phases. Or else, another possibility is that we can choose the parameters
$a$ and $e$ to be complex in such a way that $m_2$ and $m_3$ can be real.
In this later case, the Majorana phases will become zero.

Regarding the neutrino
mixing angles, we have explained in the previous section that the deviation
in $\sin\theta_{12}$ from its TBM value is undetermined at the first order level
corrections to diagonalization of the seesaw formula for neutrinos.
But now after considering
second order corrections, from Eq. (\ref{eq:r}) we can see that this deviation
can be determined in terms of $\epsilon_4$, $\epsilon_5$ and $\epsilon_6$.
In fact, out of these three $\epsilon$ parameters, only two can be determined
by solving Eqs. (\ref{eq:r}) - (\ref{eq:3rd}). We can see that by solving
Eq. (\ref{eq:e3}), we can compute $\epsilon_3$
in terms of $s_{13}$, $s$ and $\delta_{\rm CP}$. Now, by solving Eqs.
(\ref{eq:r}) $\&$ (\ref{eq:3rd}), any two of the $\epsilon_4$, $\epsilon_5$
and $\epsilon_6$ can be found in terms of the neutrino masses and mixing angles.
One among the $\epsilon_4$, $\epsilon_5$ and $\epsilon_6$ is still a free
parameter, but it should be chosen to be small in order to be consistent
with our analysis on neutrino mixing angles. After combining the results
of Eqs. (\ref{eq:res}), (\ref{eq:r}) - (\ref{eq:3rd}), we can see that
all the three neutrino mixing angles get deviations from their TBM values.
Moreover, these deviations can be fitted to experimental values by choosing
appropriate parametric space for $\epsilon_i$, which is the subject
of the next section.

\section{Numerical results}

We have explained how deviations from TBM pattern can be achieved by
introducing the $\epsilon_i$ parameters in Eq. (\ref{eq:mdd}). Here, we
numerically evaluate these parameters in order to be consistent with
the neutrino oscillation data. In this regard, in our analysis, we have
taken the best fit values for the two mass-squared differences among
the neutrinos, which are given below \cite{glo-fit}.
\begin{equation}
\Delta m_{\rm sol}^2=7.39\times 10^{-5}~{\rm eV}^2,\quad
\Delta m_{\rm atm}^2=2.525\times 10^{-3}~{\rm eV}^2.
\end{equation}
In the analysis, we have varied the three neutrino mixing angles and
the $CP$ violating Dirac
phase $\delta_{\rm CP}$ over the 3$\sigma$ ranges. These ranges are
given below \cite{glo-fit}.
\begin{eqnarray}
&&\sin^2\theta_{12}: 0.275\to0.350,\quad
\sin^2\theta_{23}: 0.418\to0.627,\quad
\sin^2\theta_{13}: 0.02045\to0.02439,
\nonumber \\
&&\delta_{\rm CP}: 125^{\rm o}\to392^{\rm o}.
\end{eqnarray}
As described in the previous section, using the allowed values for neutrino
oscillation observables, we can compute the $\epsilon_i$ parameters. Since these
parameters are complex, we have resolved them in to real and imaginary parts,
whose expressions are given below.
\begin{equation}
\epsilon_i=Re(\epsilon_i)+iIm(\epsilon_i).
\end{equation}
In order to be compatible with the above mentioned neutrino oscillation
observables, we have obtained the allowed ranges for $Re(\epsilon_i)$ and
$Im(\epsilon_i)$. These results are given in Table 1.
\begin{table}[h]
\centering
\begin{tabular}{|c|c|c|c|c|} \hline
$Re(\epsilon_1)$ & $Im(\epsilon_1)$ & $Re(\epsilon_2)$ & $Im(\epsilon_2)$,
$Im(\epsilon_3)$ & $Re(\epsilon_3)$ \\ \hline
$(-0.221,0.221)$ & $(-0.221,0.182)$ & $(-0.106,0.225)$ & $(-0.064,0.064)$ &
$(-0.15,0.095)$ \\ \hline
\end{tabular}
\begin{tabular}{|c|c|c|c|c|c|} \hline
$\phi$ & $\epsilon_4$ & $Re(\epsilon_5)$ & $Im(\epsilon_5)$ &
$Re(\epsilon_6)$ & $Im(\epsilon_6)$ \\ \hline
$0$ & $0.1$ & $(-0.084,0.462)$ & $(-0.119,0.101)$ & $(-0.375,0.168)$ &
$(-0.119,0.101)$ \\ \hline
$0$ & $-0.1$ & $(-0.282,0.26)$ & $(-0.119,0.101)$ & $(-0.175,0.367)$ &
$(-0.119,0.101)$ \\ \hline
$0$ & $0.1i$ & $(-0.182,0.362)$ & $(-0.019,0.199)$ & $(-0.275,0.267)$ &
$(-0.219,0.001)$ \\ \hline
$0$ & $-0.1i$ & $(-0.182,0.362)$ & $(-0.219,0.001)$ & $(-0.275,0.267)$ &
$(-0.019,0.199)$ \\ \hline
\end{tabular}
\caption{Allowed ranges for the real and imaginary parts of the
$\epsilon_i$ parameters. For details, see the text.}
\end{table}
In this table, the allowed ranges for real and imaginary parts of
$\epsilon_5$ and $\epsilon_6$ are obtained by fixing the values for $\epsilon_4$
and also by taking the phase $\phi=0$. Moreover, in this table,
the allowed ranges for $Im(\epsilon_2),Im(\epsilon_3)$ are same.
This result follows from Eq. (\ref{eq:res}), which implies
$Im(\epsilon_2)=Im(\epsilon_3)$.
From the results given in Table 1, we notice that except for $Re(\epsilon_5)$
and $Re(\epsilon_6)$, the magnitude of other parameters are less than
about $\sqrt{2}s_{13}\sim0.221$. The magnitudes of $Re(\epsilon_5)$
and $Re(\epsilon_6)$ can become as large as 0.46 and 0.37 respectively,
for the case of $\phi=0$ and $\epsilon_4=0.1$.
We have varied $\phi$ away from zero, by fixing $\epsilon_4=0.1$, and
have computed
real and imaginary parts of $\epsilon_5$ and $\epsilon_6$. In these cases, we
have found that the maximum values for $|Re(\epsilon_5)|$ and
$|Re(\epsilon_6)|$ to be lying between about 0.37 and 0.63, whereas,
the maximum values for $|Im(\epsilon_5)|$ and $|Im(\epsilon_6)|$ are found
to be less than 0.2.

We have explained previously
that the diagonlization of the seesaw formula of our model is done by
assuming that the magnitudes of real and imaginary parts of $\epsilon_i$ to
be less than or of the order of
$\sqrt{\frac{\Delta m^2_{\rm sol}}{\Delta m^2_{\rm atm}}}\sim\sin\theta_{13}$.
From the numerical results presented above, we can notice that,
real and imaginary parts of $\epsilon_i$ satisfy the above mentioned
assumption, except for $Re(\epsilon_5)$ and $Re(\epsilon_6)$. The maximum
values for $|Re(\epsilon_5)|$ and $|Re(\epsilon_6)|$ can be about 0.4,
depending on $\epsilon_4$ and $\phi$ values. At these maximum values,
the analytic expressions presented in previous sections may not give
accurate results, since square of 0.4 is not negligible in comparison
to unity. One can notice that $\epsilon_5$ and $\epsilon_6$ contribute
linearly to neutrino oscillation observables in the second order corrections
of our analysis. As a result of this,
corrections at the third order level to the above mentioned observables
are not negligible
around these maximum values. On the other hand, one would like to know,
if by restricting the values of $|Re(\epsilon_5)|$ and $|Re(\epsilon_6)|$
to be small, the analytic expressions of previous sections are
sufficient enough to give accurate numerical results. For this reason,
we have computed allowed values for neutrino mixing angles and
$\delta_{\rm CP}$ by demanding $|Re(\epsilon_5)|$ and $|Re(\epsilon_6)|$
to be less than 0.23 for the case of $\phi=0$ and $\epsilon_4=0.1$.
These results are given in Figure 1.
\begin{figure}[!h]
\begin{center}

\includegraphics[height=3.0in,width=3.0in]{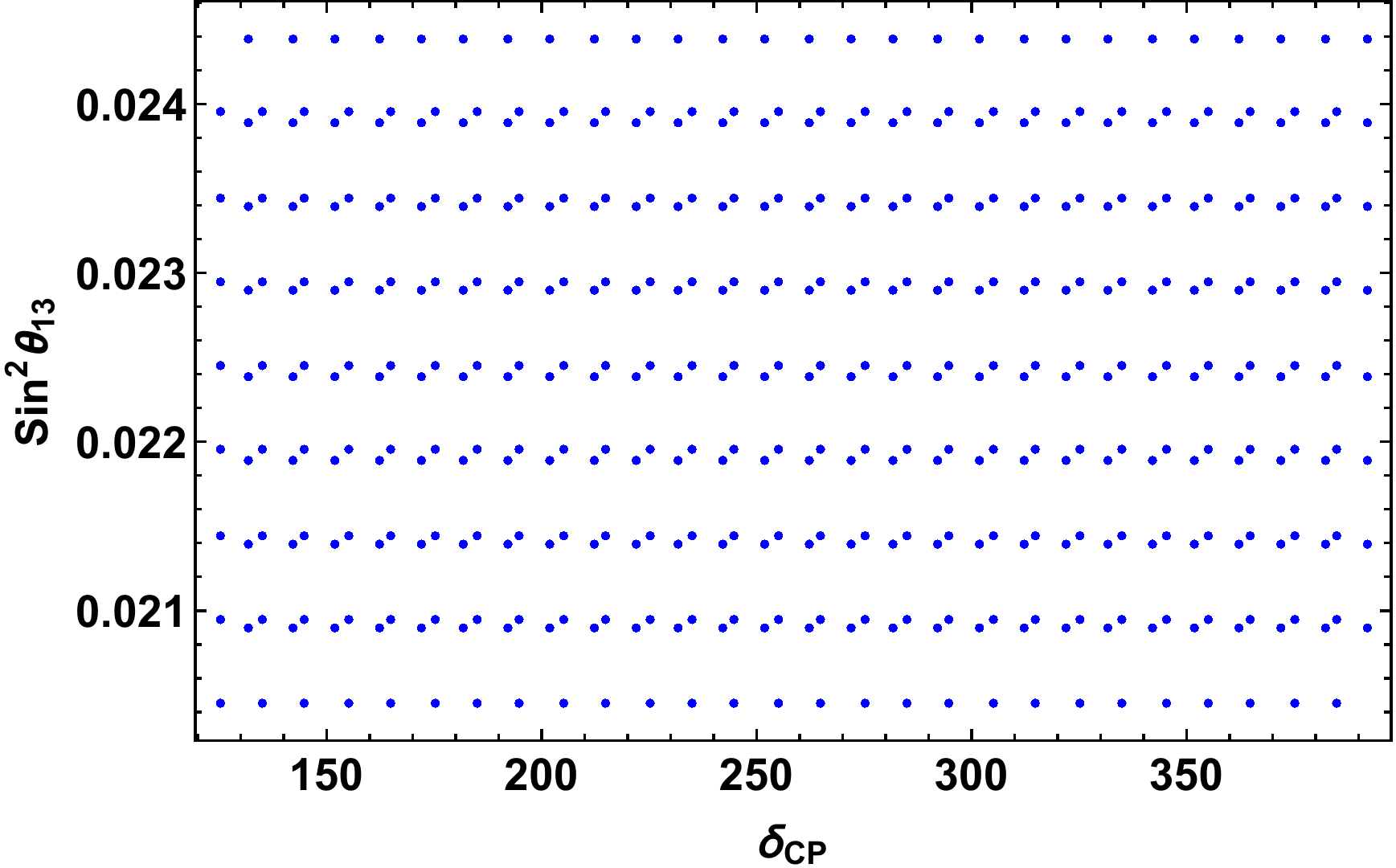}
\includegraphics[height=3.0in,width=3.0in]{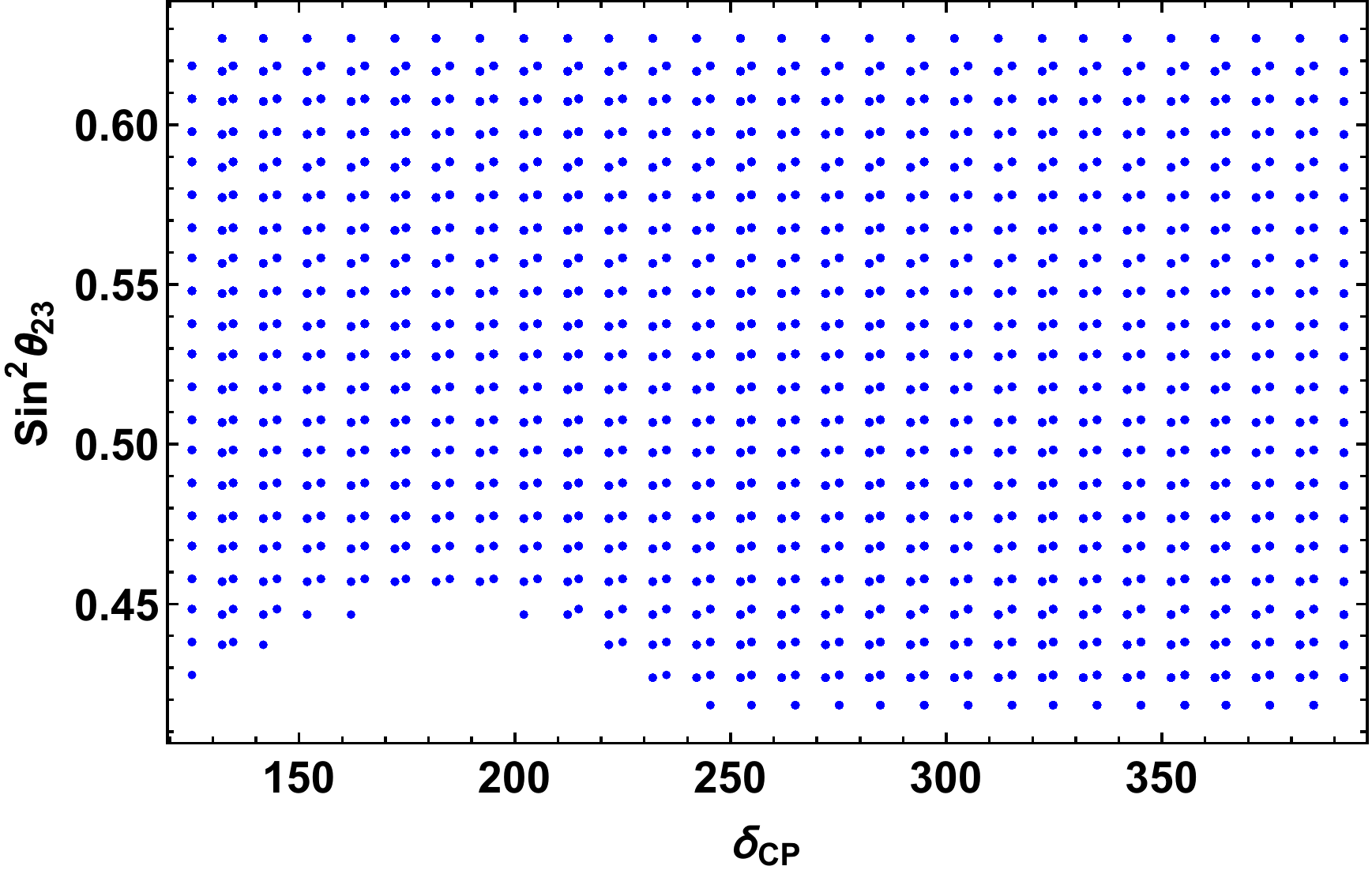}

\includegraphics[height=3.0in,width=3.0in]{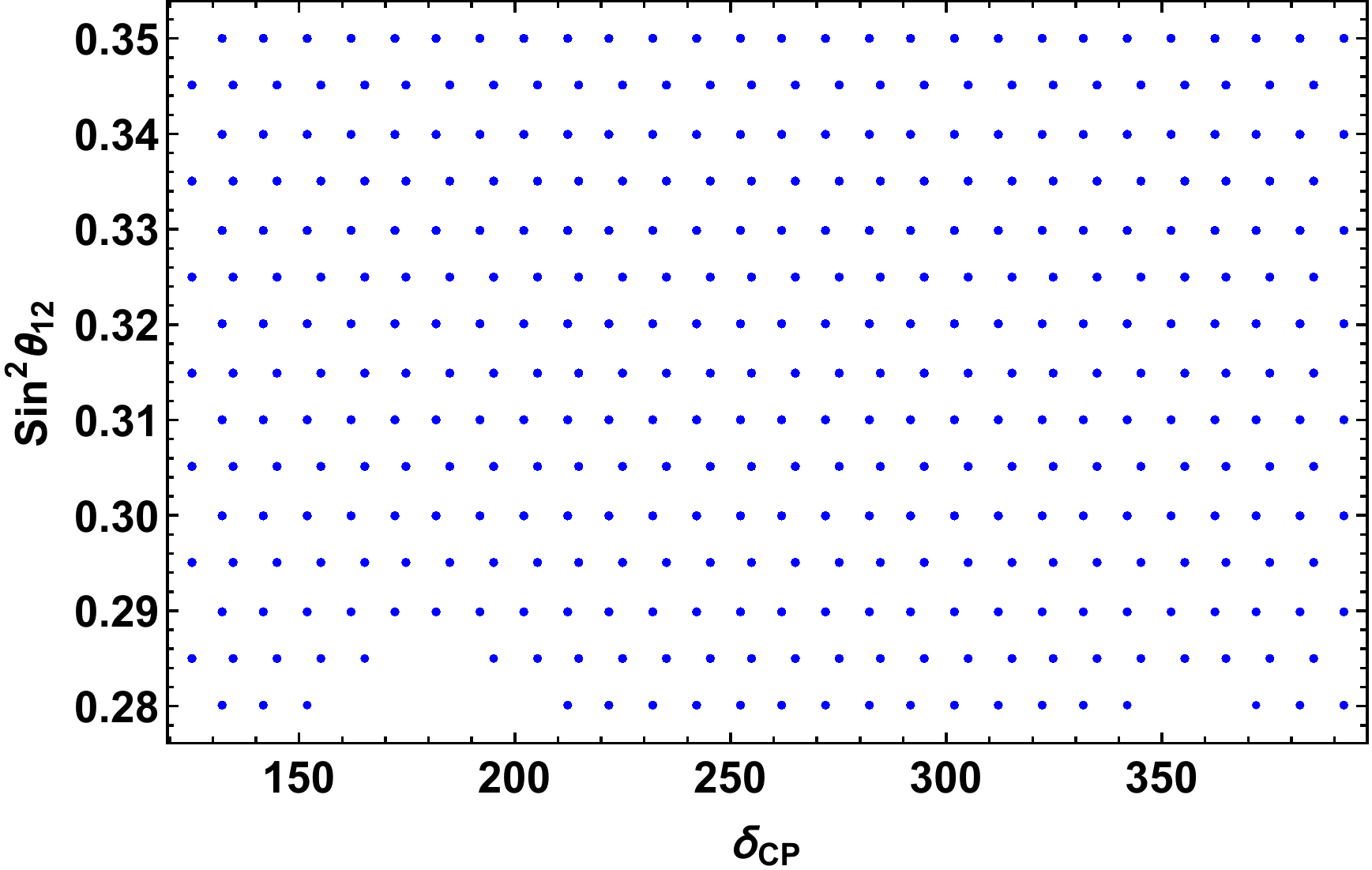}

\end{center}
\caption{Allowed regions in neutrino mixing angles and $\delta_{\rm CP}$
by demanding $|Re(\epsilon_5)|$ and $|Re(\epsilon_6)|$ to be less than 0.23,
for the case of $\phi=0$ and $\epsilon_4=0.1$.
$\delta_{\rm CP}$ is expressed in degrees.}
\end{figure}
Now, after combining the results of Table 1, one can notice that in the allowed
regions of Figure 1, the magnitudes of real and imaginary parts of all
$\epsilon_i$ are less than 0.23, which is about $\sqrt{2}s_{13}$. Hence,
in the allowed regions of Figure 1, $\epsilon_i$ satisfy the assumptions
we have made in order to diagonalize the seesaw formula of our model.

We have discretized the axes of $s_{13}^2,s_{23}^2,s_{12}^2,\delta_{\rm CP}$
in order to obtain the results in Figure 1. From this figure, one can notice
that
the allowed range for $s_{13}^2$ is almost unconstrained. However, the
allowed regions for $s_{23}^2$ and $s_{12}^2$ are constrained for
some specific values of $\delta_{\rm CP}$. A notable feature from Figure 1
is that points with $s_{12}^2=0.275$ are excluded
for any value of $\delta_{\rm CP}$. Although the results in Figure 1
are obtained for $\phi=0$ and $\epsilon_4=0.1$, a similar analysis can
be done for any other values of $\phi$ and $\epsilon_4$. Hence, in our
work, we can find a region where $\epsilon_i$ parameters are small and
consistently explain the deviation from TBM pattern in the neutrino sector.

Recently, results from global fits to neutrino oscillation data have been
updated in ref.\cite{upglo}. These results prefer normal hierarchy for
neutrino masses. In our framework, which is based on CSD, the analytic
results from previous section show that neutrino masses have normal hierarchy.
Hence, the prediction of our CSD scenario is favourable by the recent
neutrino oscillation data. In the case of normal mass hierarchy, recent
results from neutrino oscillation data prefer second octant for
$\sin^2\theta_{23}$ with the best fit value of $0.566$ \cite{upglo}. The
CP violating phase $\delta_{\rm CP}$, in the above case, has a best fit value of
$1.2\pi$ \cite{upglo}. From the results given in Figure 1, we notice
lower octant
values for $\sin^2\theta_{23}$ are excluded for $\delta_{\rm CP}$ around
$\pi$. Hence, the numerical results of our work are compatible with the
recent neutrino oscillation data.

\section{A model for our Dirac mass matrix}

In this section, we construct a model in order to justify the structure
of our Dirac mass matrix of Eq. (\ref{eq:mdd}) and also
explain the small values for $\epsilon_i$. For this purpose,
we introduce a flavor symmetry $SU(3)$ and also
the following scalar fields: $\phi_{a},\phi_{s},
\phi^\prime_{a},\phi^\prime_{s}$. These scalar fields are singlets
under the standard model gauge group, but otherwise, charged under the
$SU(3)$. The lepton doublets $L$, where
we have suppressed generation index, are charged under this flavor symmetry.
The Higgs doublet $H$ and the two right-handed neutrinos $\nu_R^{atm}$,
$\nu_R^{sol}$ are singlets under $SU(3)$. To get the masses for
right-handed neutrinos, we introduce the following additional scalar fields,
which are standard model gauge singlets: $\chi_a$, $\chi_s$.
To explain the smallness of $\epsilon_i$ parameters, we propose the
scalar field $\xi$ which is a standard model gauge singlet.
To forbid unwanted interactions
in our model we introduce a discreet symmetry $Z_3\times Z^\prime_3$.
In Table 2,
charges assignments of the fields, which are relevant to neutrino sector,
are given.
\begin{table}[h]
\centering
\begin{tabular}{|c|c|c|c|c|c|c|c|c|c|c|c|}\hline
 & $\phi_{a}$ & $\phi_{s}$ & $\phi^\prime_{a}$ & $\phi^\prime_{s}$ &
$\xi$ & $\chi_a$ & $\chi_s$ & $\nu_R^{atm}$ & $\nu_R^{sol}$ & $L$
& $H$ \\ \hline
$SU(3)$ & 3 & 3 & 3 & 3 & 1 & 1 & 1 & 1 & 1 & 3 & 1 \\ \hline
$Z_3$ & $\omega$ & $\omega^2$ & $\omega$ & $\omega^2$ & 1 & $\omega^2$ &
$\omega$ & $\omega^2$ & $\omega$ & 1 & 1 \\ \hline
$Z^\prime_3$ & $\omega^2$ & $\omega^2$ & $\omega$ & $\omega$ & $\omega$ &
$\omega$ & $\omega$ & $\omega$ & $\omega$ & 1 & 1 \\ \hline
\end{tabular}
\caption{Charge assignments of the relevant fields
under the flavor symmetry $SU(3)\times Z_3\times Z^\prime_3$ are given.
Here, $\omega=e^{2\pi i/3}$. For other details, see the text.}
\end{table}
With these charge assignments, the leading terms in the Lagrangian are
\begin{eqnarray}
{\cal L} &=& \frac{\phi_{a}}{M_P}\bar{L}\nu_R^{atm}H+
\frac{\phi_{s}}{M_P}\bar{L}\nu_R^{sol}H+
\frac{\xi}{M_P}\frac{\phi^\prime_{a}}{M_P}\bar{L}\nu_R^{atm}H+
\frac{\xi}{M_P}\frac{\phi^\prime_{s}}{M_P}\bar{L}\nu_R^{sol}H
\nonumber \\
&& +\frac{\chi_a}{2}
\overline{(\nu_R^{atm})^c}\nu_R^{atm}+\frac{\chi_s}{2}
\overline{(\nu_R^{sol})^c}\nu_R^{sol}+{h.c.}
\label{eq:lag}
\end{eqnarray}
Here, $M_P\sim 2\times 10^{18}$ GeV is the reduced Planck scale, which is the
cutoff scale for this model. The reason for choosing the Planck scale
as the cutoff of the model is explained later this section.

The first four terms of Eq. (\ref{eq:lag}) generate effective Yukawa
couplings for neutrinos after the following scalar fields acquire
vevs: $\phi_{a},\phi_{s},\phi^\prime_{a},\phi^\prime_{s},\xi$.
The vevs of $\phi_{a},\phi_{s}$ give leading contribution to
effective Yukawa couplings.
In order to explain the structure of Dirac mass matrix of Eq.
(\ref{eq:mdd}) we assume that these vevs to have the following pattern
\begin{equation}
\frac{\langle\phi_{a}\rangle}{M_P}=y_a\left(\begin{array}{c}
0\\1\\1 \end{array}\right),\quad
\frac{\langle\phi_{s}\rangle}{M_P}=y_s\left(\begin{array}{c}
1\\1\\-1 \end{array}\right)
\label{eq:vev1}
\end{equation}
Here, $y_a,y_s$ are dimensionless quantities. The
above pattern of vevs can be obtained by tuning parameters in
the scalar potential of this model, which is described in the next section.
It is to be noted that the problem
related to vev pattern
of Eq. (\ref{eq:vev1}) has been addressed in ref.\cite{csd}. The
vevs of $\phi^\prime_{a},\phi^\prime_{s},\xi$ give sub-leading
contribution to Yukawa couplings for neutrinos. Here, we need not assume
any pattern for the vevs of $\phi^\prime_{a},\phi^\prime_{s}$.
Hence, after writing $\frac{\langle\xi\rangle}{M_P}=\epsilon$,
we can have
\begin{equation}
\frac{\langle\xi\rangle}{M_P}\frac{\langle\phi_{a}^\prime\rangle}
{M_P}=y_a\left(\begin{array}{c} y_1\\y_2\\y_3 \end{array}\right)
\epsilon = y_a\left(\begin{array}{c} \epsilon_1\\\epsilon_2\\\epsilon_3
\end{array}\right),\quad
\frac{\langle\xi\rangle}{M_P}\frac{\langle\phi_{s}^\prime\rangle}
{M_P}=y_s\left(\begin{array}{c} y^\prime_1\\y^\prime_2\\y^\prime_3
\end{array}\right)\epsilon =
y_s\left(\begin{array}{c} \epsilon_4\\\epsilon_5\\\epsilon_6
\end{array}\right)
\label{eq:vev2}
\end{equation}
Here, $y_i,y^\prime_i$, where $i=1,\cdots,3$, are ${\cal O}(1)$ parameters.
By taking
$\frac{\langle\xi\rangle}{M_P}=\epsilon\sim 0.1$, we get
all $\epsilon_i$ to be around 0.1. We can see that the smallness of
$\epsilon_i$ parameters can be explained if the field $\xi$ acquire
vev around one order less than the $M_P$. Now, using Eqs.
(\ref{eq:vev1}) $\&$ (\ref{eq:vev2}) in Eq. (\ref{eq:lag}), after
electroweak symmetry breaking, we get
the structure of Dirac mass matrix which is proposed in Eq. (\ref{eq:mdd}).
Finally, the last two terms of Eq. (\ref{eq:lag}) give diagonal masses to
right-handed neutrinos, after the fields $\chi_a,\chi_s$ acquire vevs.

In the above, by proposing a model, we have explained the mass structures of
Dirac and right-handed neutrinos of this work. In order to explain these mass
structures, the extra scalar fields proposed in this model need to acquire
vevs and thereby break the flavor symmetry $SU(3)\times Z_3\times Z^\prime_3$
spontaneously.
Here we quantify the scales of these vevs. It is stated above that
$\langle\chi_a\rangle$ and $\langle\chi_s\rangle$ generate masses
for right-handed neutrinos. Requiring that these masses to be around
1 TeV, we should have: $\langle\chi_a\rangle,\langle\chi_s\rangle
\sim$ 1 TeV. One motivation for choosing TeV scale masses for right-handed
neutrinos is that they can be detected in the LHC experiment.
Another motivation for choosing
the above mass scale for right-handed neutrinos is shortly explained below. 
The vev of $\xi$ can be
found from the fact that it explains the
smallness of $\epsilon_i$ parameters. In order to explain this smallness,
we have described above that we need to have $\frac{\langle\xi\rangle}{M_P}
\sim 0.1$. From this we get $\langle\xi\rangle\sim 10^{17}$ GeV. Finally,
the vevs of $\phi_{a},\phi_{s},\phi^\prime_{a},\phi^\prime_{s}$
can be determined from the reasoning that they generate effective Yukawa
couplings for neutrinos. Since we have taken right-handed neutrino masses to be
around 1 TeV, from seesaw formula for active neutrinos, we can estimate
the magnitude of Yukawa couplings for neutrinos by having the active neutrino
masses to be of ${\cal O}(0.1)$ eV. From this estimation, we have found that
$\langle\phi_{a}\rangle,\langle\phi_{s}\rangle,\langle\phi^\prime_{a}
\rangle,\langle\phi^\prime_{s}\rangle\sim 10^{12}$ GeV. After finding
the vevs of the additional scalar fields of this model, we notice that
there is a large hierarchy among these vevs. We can achieve this hierarchy,
in this model, by appropriately fixing the relevant parameters in
the scalar potential among the above mentioned scalar fields.
Analysis related to the scalar potential of our model is presented in
Appendix B.

It is stated that $M_P$ is the cutoff scale for the above described model.
One can notice that the hierarchy between $\langle\xi\rangle$ and
$M_P$ is very less. However, the hierarchy in the vevs of other scalar
fields such as $\phi_{a},\phi_{s},\phi^\prime_{a},\phi^\prime_{s},
\chi_a,\chi_s$ is very large with respect to $M_P$. Here, we explain
this hierarchy
by motivating the above described model from supersymmetry \cite{susy}.
Since supersymmetry is not exact symmetry, one possibility is to break
supersymmetry spontaneously by hidden sector fields and the effects of
this breaking are mediated to visible sector via gravity mediated
interactions \cite{susy}. Models based on this mechanism are known as
supergravity models, where hidden sector fields can interact with visible
sector fields with Planck scale suppressed non-renormalizable terms.
In these models, hidden sector fields can acquire vevs around the intermediate
scale of $\Lambda\sim10^{11}$ GeV and the TeV scale can be generated by
$\frac{\Lambda^2}{M_P}$. Based on this, models have been proposed in order
to conceive TeV scale masses for right-handed neutrinos \cite{tevrh}.
In our model, which is described above, the vevs for $\chi_a,\chi_s$
are around TeV and the vevs for $\phi_{a},\phi_{s},\phi^\prime_{a},
\phi^\prime_{s}$ are close to the intermediate scale. Hence, we can explain
the hierarchy in the vev of these fields by embedding our model in
a supergravity setup. In fact, for this reason we have chosen $M_P$ as
the cutoff scale to our model.

Above, we have motivated the hierarchy in the vevs of scalar fields of our
model from a supergravity setup. On the other hand, in order to achieve this
hierarchy
in the current framework, we have carried out an analysis on the scalar
potential of our model in Appendix B. In this appendix, we have given
the invariant scalar potential under the flavor symmetry
$SU(3)\times Z_3\times Z_3^\prime$.
After minimizing
this scalar potential, the vevs of various scalar fields of our model have
been determined. It is shown in Appendix B that by tuning the parameters
of the scalar potential, the required hierarchy among the vevs of the
scalar fields can be achieved. It has been argued that the desired vacuum
alignment for $\phi_a$ and $\phi_s$ can be achieved by tuning the necessary
parameters in the scalar potential. Although the scalar potential in Appendix
B is at tree level, due to large number of parameters, the above mentioned
vacuum alignment is still possible even after including the radiative
corrections to the scalar potential. It is described in Appendix B that
in order to achieve the desired hierarchy in the vevs of scalar fields, some
couplings in the scalar potential should be suppressed to as low as
$10^{-32}$. To explain the smallness of these couplings, one can extend
the flavor symmetry of our model to $SU(3)\times Z_3\times Z_3^\prime\times
Z_2\times Z_2^\prime$. We have noticed that under the additional symmetry
$Z_2\times Z_2^\prime$, charge assignments for fields can be done in such a
way that terms in Eq. (\ref{eq:lag}) are allowed but the terms in the scalar
potential with couplings of the order of $10^{-32}$ are forbidden.
After doing that, one can
motivate the smallness of these couplings as a soft breaking of the additional
flavor symmetry $Z_2\times Z_2^\prime$. To explain the smallness of other
couplings in the scalar potential, either one can extend the above flavor
symmetry or one needs to device a new mechanism.

It is described previously that apart from the Higgs field, rest of the
scalar fields are charged under the flavor symmetry $SU(3)\times
Z_3\times Z_3^\prime$. This symmetry is spontaneously broken by the vevs
of the scalar fields. Since these fields are complex, apart from the
Higgs boson, a total of thirty real scalar fields exist in our model. By
choosing the flavor symmetry $SU(3)$ to be gauged, after spontaneous symmetry
breaking, apart from the Higgs boson,
twenty two real scalar fields remain in our model. All these fields have mixing
masses. We estimate the masses for these fields to be around
the scales at which they acquire vevs. The gauge bosons of the flavor
symmetry $SU(3)$ can
get masses around $10^{12}$ GeV. Since lepton doublets are charged under
the flavor $SU(3)$, the above gauge bosons have couplings to leptons.
A study on the
phenomenology of the additional fields of our model is out of the scope of
this work.

\section{Conclusions}

In this work, we have attempted to explain the neutrino mixing in order
to be consistent with the current neutrino oscillation data.
From the current data,
it is known that $\theta_{13}\neq0$, and hence, the neutrino
mixing angles deviate away from the TBM pattern. Earlier, to explain the
TBM pattern in neutrino sector, CSD model has been proposed. Here, we
have considered a phenomenological model, where we have modified
the neutrino Yukawa couplings of CSD model, by introducing
small $\epsilon_i$ parameters which are complex. We have assumed real
and imaginary parts of $\epsilon_i$ to be less than or of the order of
$\sin\theta_{13}\sim\sqrt{\frac{\Delta m^2_{\rm sol}}{\Delta m^2_{\rm atm}}}$.
Thereafter, we have followed an approximation procedure in order to diagonalize
the seesaw formula for light neutrinos in our model. We have computed
expressions, up to second order level, to neutrino masses and mixing angles
in terms of small $\epsilon_i$ parameters. Using these expressions we have
demonstrated that neutrino mixing angles can deviate away from their TBM values
by appropriately choosing the $\epsilon_i$ values. Finally, we have constructed
a model in order to justify the neutrino Yukawa coupling structure of our
phenomenological model.

\section*{Appendix A}

The full form of second order terms in
$\frac{1}{\sqrt{\Delta m_{\rm atm}^2}}m_\nu^d$ are given below.
\begin{eqnarray}
&&\frac{1}{\sqrt{\Delta m_{\rm atm}^2}}m_{\nu(2)}^d=
\frac{1}{\sqrt{\Delta m_{\rm atm}^2}}\left(
\begin{array}{ccc}
x^{\prime\prime}_{11} & x^{\prime\prime}_{12} & x^{\prime\prime}_{13} \\
x^{\prime\prime}_{12} & x^{\prime\prime}_{22} & x^{\prime\prime}_{23} \\
x^{\prime\prime}_{13} & x^{\prime\prime}_{23} & x^{\prime\prime}_{33}
\end{array}\right),
\nonumber \\
x^{\prime\prime}_{11}&=&\frac{e^2}{6M_{\rm atm}}[4\epsilon_1^2+\epsilon_2^2
+(\epsilon_3+2s)^2-4\sqrt{2}e^{i\delta_{\rm CP}}s_{13}(\epsilon_3+2s)+
8e^{2i\delta_{\rm CP}}s^2_{13}
\nonumber \\
&&-2\epsilon_2(\epsilon_3+2s-2\sqrt{2}
e^{i\delta_{\rm CP}}s_{13})-4\epsilon_1(\epsilon_2-\epsilon_3-2s+
2\sqrt{2}e^{i\delta_{\rm CP}}s_{13})],
\nonumber \\
x^{\prime\prime}_{12}&=&\frac{a^2}{\sqrt{2}M_{\rm sol}}(2\epsilon_4-
\epsilon_5+\epsilon_6-3r)+\frac{e^2}{6M_{\rm atm}}[\sqrt{2}(2\epsilon_1^2
-\epsilon_2^2)-\sqrt{2}(\epsilon_3+2s)^2
\nonumber \\
&&+2e^{i\delta_{\rm CP}}s_{13}
(\epsilon_3+2s)+4\sqrt{2}e^{2i\delta_{\rm CP}}s^2_{13}+\epsilon_1(
\sqrt{2}(\epsilon_2-\epsilon_3-2s)-8e^{i\delta_{\rm CP}}s_{13})
\nonumber \\
&&+2\epsilon_2(\sqrt{2}(\epsilon_3+2s)-e^{i\delta_{\rm CP}}s_{13})],
\nonumber \\
x^{\prime\prime}_{13}&=&\frac{e^2}{\sqrt{2}M_{\rm sol}}[-\epsilon_2^2
+\epsilon_3^2+2\epsilon_1(\epsilon_2+\epsilon_3-r)+2\epsilon_3r+4\epsilon_3s
+4rs+2s^2-4\sqrt{2}e^{i\delta_{\rm CP}}s_{13}\epsilon_3
\nonumber \\
&&+2\sqrt{2}e^{i\delta_{\rm CP}}s_{13}r-2\epsilon_2(r-2s+2\sqrt{2}
e^{i\delta_{\rm CP}}s_{13})],
\nonumber \\
x^{\prime\prime}_{22}&=&\frac{2a^2}{M_{\rm sol}}(\epsilon_4+\epsilon_5
-\epsilon_6)+\frac{e^2}{3M_{\rm atm}}[\epsilon_1^2+\epsilon_2^2+
(\epsilon_3+2s)^2+2\sqrt{2}e^{i\delta_{\rm CP}}s_{13}(\epsilon_3+2s)
\nonumber \\
&&+2e^{2i\delta_{\rm CP}}s^2_{13}+2\epsilon_1(\epsilon_2-\epsilon_3-2s
-\sqrt{2}e^{i\delta_{\rm CP}}s_{13})-2\epsilon_2(\epsilon_3+2s
+\sqrt{2}e^{i\delta_{\rm CP}}s_{13})],
\nonumber \\
x^{\prime\prime}_{23}&=&\frac{\sqrt{3}a^2}{2M_{\rm sol}}[\sqrt{2}
(\epsilon_5+\epsilon_6+2s)+2e^{-i\delta_{\rm CP}}s_{13}]+
\frac{e^2}{2\sqrt{3}M_{\rm atm}}[\sqrt{2}(\epsilon_2^2-\epsilon_3^2+
\epsilon_3r)
\nonumber \\
&&+\sqrt{2}\epsilon_1(\epsilon_2+\epsilon_3+2r)-\sqrt{2}(4\epsilon_3s-2rs
+2s^2)-4e^{i\delta_{\rm CP}}s_{13}(\epsilon_3+r)
\nonumber \\
&&-\epsilon_2(\sqrt{2}r+4\sqrt{2}s+4e^{i\delta_{\rm CP}}s_{13})],
\nonumber \\
x^{\prime\prime}_{33}&=&\frac{e^2}{2M_{\rm atm}}
[4\sqrt{2}\epsilon_1s_{13}e^{-i\delta_{\rm CP}}+\epsilon_2^2+\epsilon_3^2
-4\epsilon_3s+2\epsilon_2(\epsilon_3+2s)-4(s^2+s_{13}^2)].
\label{eq:2ndrel}
\end{eqnarray}

\section*{Appendix B}

Here we analyze the scalar potential of the model, which is
described in section 6. The invariant scalar potential of this
model is given below.
\begin{eqnarray}
V&=&m_{\phi_a}^2\phi_a^\dagger\phi_a+m_{\phi_s}^2\phi_s^\dagger\phi_s
+m_{\phi_a^\prime}^2\phi_a^{\prime\dagger}\phi_a^\prime+m_{\phi_s^\prime}^2
\phi_s^{\prime\dagger}\phi_s^\prime+m_{\chi_a}^2\chi_a^*\chi_a
+m_{\chi_s}^2\chi_s^*\chi_s+m_\xi^2\xi^*\xi
\nonumber \\
&&+m_H^2H^\dagger H+\lambda_{\phi_a}(\phi_a^\dagger\phi_a)^2+\lambda_{\phi_s}
(\phi_s^\dagger\phi_s)^2
+\lambda_{\phi_a^\prime}(\phi_a^{\prime\dagger}\phi_a^\prime)^2
+\lambda_{\phi_s^\prime}(\phi_s^{\prime\dagger}\phi_s^\prime)^2
+\lambda_{\chi_a}(\chi_a^*\chi_a)^2
\nonumber \\
&&+\lambda_{\chi_s}(\chi_s^*\chi_s)^2+\lambda_{\xi}(\xi^*\xi)^2
+\lambda_{H}(H^\dagger H)^2
+\lambda_{\phi_aH}(\phi_a^\dagger\phi_a)(H^\dagger H)
+\lambda_{\phi_sH}(\phi_s^\dagger\phi_s)(H^\dagger H)
\nonumber \\
&&+\lambda_{\phi^\prime_aH}(\phi_a^{\prime\dagger}\phi_a^\prime)(H^\dagger H)
+\lambda_{\phi_s^\prime H}(\phi_s^{\prime\dagger}\phi_s^\prime)(H^\dagger H)
+\lambda_{\chi_aH}(\chi_a^*\chi_a)(H^\dagger H)
+\lambda_{\chi_sH}(\chi_s^*\chi_s)(H^\dagger H)
\nonumber \\
&&+\lambda_{\xi H}(\xi^*\xi)(H^\dagger H)
+\left[a_1(\phi_a^\dagger\phi_a^\prime)\xi
+a_2(\phi_a^\dagger\phi_s^\prime)\chi_a
+a_3(\phi_s^\dagger\phi_a^\prime)\chi_s+a_4(\phi_s^\dagger\phi_s^\prime)\xi
+a_5\chi_a\chi_s\xi\right.
\nonumber \\
&&\left. +a_6\xi^3+a_7\chi_a^3+a_8\chi_s^3+h.c.\right]
+b_1(\phi_a^\dagger\phi_a)(\phi_s^\dagger\phi_s)
+b_2(\phi_a^\dagger\phi_a)(\phi_a^{\prime\dagger}\phi_a^\prime)
+b_3(\phi_a^\dagger\phi_a)(\phi_s^{\prime\dagger}\phi_s^\prime)
\nonumber \\
&&+b_4(\phi_s^\dagger\phi_s)(\phi_a^{\prime\dagger}\phi_a^\prime)
+b_5(\phi_s^\dagger\phi_s)(\phi_s^{\prime\dagger}\phi_s^\prime)
+b_6(\phi_a^{\prime\dagger}\phi_a^\prime)(\phi_s^{\prime\dagger}\phi_s^\prime)
+b_7(\phi_a^\dagger\phi_s^\prime)(\phi_s^{\prime\dagger}\phi_a)
\nonumber \\
&&+b_8(\phi_a^\dagger\phi_s)(\phi_s^\dagger\phi_a)
+\left[b_9(\phi_s^\dagger\phi_a)(\phi_a^{\prime\dagger}\phi_s^\prime)
+b_{10}(\phi_s^\dagger\phi_s^\prime)(\phi_a^{\prime\dagger}\phi_a)+h.c.\right]
+b_{11}(\phi_s^\dagger\phi_a^\prime)(\phi_a^{\prime\dagger}\phi_s)
\nonumber \\
&&+b_{12}(\phi_s^{\prime\dagger}\phi_s)(\phi_s^\dagger\phi_s^\prime)
+b_{13}(\phi_s^{\prime\dagger}\phi_a^\prime)(\phi_a^{\prime\dagger}
\phi_s^\prime)
+b_{14}(\phi_a^\dagger\phi_a^\prime)(\phi_a^{\prime\dagger}\phi_a)
+c_1(\phi_a^\dagger\phi_a)\xi^*\xi+c_2(\phi_s^\dagger\phi_s)\xi^*\xi
\nonumber \\
&&+c_3(\phi_a^{\prime\dagger}\phi_a^\prime)\xi^*\xi
+c_4(\phi_s^{\prime\dagger}\phi_s^\prime)\xi^*\xi
+c_5(\phi_a^\dagger\phi_a)\chi_a^*\chi_a
+c_6(\phi_s^\dagger\phi_s)\chi_a^*\chi_a
+c_7(\phi_a^{\prime\dagger}\phi_a^\prime)\chi_a^*\chi_a
\nonumber \\
&&+c_8(\phi_s^{\prime\dagger}\phi_s^\prime)\chi_a^*\chi_a
+c_9(\phi_a^\dagger\phi_a)\chi_s^*\chi_s
+c_{10}(\phi_s^\dagger\phi_s)\chi_s^*\chi_s
+c_{11}(\phi_a^{\prime\dagger}\phi_a^\prime)\chi_s^*\chi_s
+c_{12}(\phi_s^{\prime\dagger}\phi_s^\prime)\chi_s^*\chi_s
\nonumber \\
&&+\left[c_{13}(\phi_a^\dagger\phi_s)\chi_a^*\chi_s
+c_{14}(\phi_s^\dagger\phi_a)\xi^*\chi_s
+c_{15}(\phi_a^\dagger\phi_s)\xi^*\chi_a
+c_{16}(\phi_a^\dagger\phi_a^\prime)\chi_a^*\chi_s^*
+c_{17}(\phi_s^{\prime\dagger}\phi_a)\chi_s\xi\right.
\nonumber \\
&&\left.+c_{18}(\phi_s^\dagger\phi_a^\prime)\chi_a^*\xi^*
+c_{19}(\phi_a^{\prime\dagger}\phi_s)\chi_s\chi_s
+c_{20}(\phi_s^{\prime\dagger}\phi_s)\chi_a\chi_s
+c_{21}(\phi_a^{\prime\dagger}\phi_s^\prime)\xi^*\chi_a
+c_{22}(\phi_a^{\prime\dagger}\phi_s^\prime)\chi_a^*\chi_s\right.
\nonumber \\
&&\left.+c_{23}(\phi_a^{\prime\dagger}\phi_s^\prime)\chi_s^*\xi
+c_{24}(\phi_s^{\prime\dagger}\phi_a)\chi_a\chi_a
+c_{25}(\phi_a^\dagger\phi_a^\prime)\xi^*\xi^*
+c_{26}(\phi_s^\dagger\phi_s^\prime)\xi^*\xi^*+h.c.\right]
\nonumber \\
&&+d_1\xi^*\xi\chi_a^*\chi_a+d_2\xi^*\xi\chi_s^*\chi_s
+d_3\chi_s^*\chi_s\chi_a^*\chi_a+\left[d_4\xi^*\xi^*\chi_a\chi_s
+d_5\xi^*\chi_a^*\chi_s\chi_s+d_6\xi^*\chi_s^*\chi_a\chi_a\right.
\nonumber \\
&&\left.+h.c.\right]
\label{eq:pot}
\end{eqnarray}
In the above equation, the parameters in the first eight terms have
mass-squared dimensions and the $a$ parameters have dimensions of mass.
Rest of the parameters in Eq. (\ref{eq:pot}) are dimensionless.
After minimizing the above scalar potential, different scalar fields of
the model acquire vevs, which satisfy the following relations.
\begin{eqnarray}
&& m_H^2+2\lambda_H\langle H^\dagger H\rangle+\lambda_{\phi_aH}\langle
\phi_a^\dagger\phi_a\rangle+\lambda_{\phi_sH}\langle\phi_s^\dagger\phi_s\rangle
+\lambda_{\phi_a^\prime H}\langle\phi_a^{\prime\dagger}\phi_a^\prime\rangle
+\lambda_{\phi_s^\prime H}\langle\phi_s^{\prime\dagger}\phi_s^\prime\rangle
\nonumber \\
&& +\lambda_{\chi_aH}\langle\chi_a^*\chi_a\rangle+
\lambda_{\chi_sH}\langle\chi_s^*\chi_s\rangle
+\lambda_{\xi H}\langle\xi^*\xi\rangle =0,
\label{eq:h}
\\
&& \left[m_{\chi_a}^2+2\lambda_{\chi_a}\langle\chi_a^*\chi_a\rangle+
c_5\langle\phi_a^\dagger\phi_a\rangle+c_6\langle\phi_s^\dagger\phi_s\rangle
+c_7\langle\phi_a^{\prime\dagger}\phi_a^\prime\rangle
+c_8\langle\phi_s^{\prime\dagger}\phi_s^\prime\rangle
+d_3\langle\chi_s^*\chi_s\rangle+d_1\langle\xi^*\xi\rangle\right.
\nonumber \\
&&\left.+\lambda_{\chi_aH}\langle H^\dagger H\rangle\right]
\langle\chi_a^*\rangle
+a_2\langle\phi_a^\dagger\phi_s^\prime\rangle
+\left[c_{13}^*\langle\phi_s^\dagger\phi_a\rangle
+c_{22}^*\langle\phi_s^{\prime\dagger}\phi_a^\prime\rangle
+d_5^*\langle\xi\chi_s^*\rangle\right]\langle\chi_s^*\rangle
+\left[c_{15}\langle\phi_a^\dagger\phi_s\rangle\right.
\nonumber \\
&& \left.+c_{21}\langle\phi_a^{\prime\dagger}\phi_s^\prime\rangle\right]
\langle\xi^*\rangle
+c_{18}^*\langle\phi_a^{\prime\dagger}\phi_s\rangle\langle\xi\rangle
+\left[a_5\langle\xi\rangle+c_{16}^*\langle\phi_a^{\prime\dagger}\phi_a\rangle
+c_{20}\langle\phi_s^{\prime\dagger}\phi_s\rangle
+d_4\langle\xi^*\xi^*\rangle\right]\langle\chi_s\rangle
\nonumber \\
&&+\left[2c_{24}\langle\phi_s^{\prime\dagger}\phi_a\rangle
+2d_6\langle\xi^*\chi_s^*\rangle\right]\langle\chi_a\rangle=0,
\label{eq:chia}
\\
&&\left[m_{\chi_s}^2+2\lambda_{\chi_s}\langle\chi_s^*\chi_s\rangle
+c_9\langle\phi_a^\dagger\phi_a\rangle+c_{10}\langle\phi_s^\dagger\phi_s\rangle
+c_{11}\langle\phi_a^{\prime\dagger}\phi_a^\prime\rangle
+c_{12}\langle\phi_s^{\prime\dagger}\phi_s^\prime\rangle
+d_2\langle\xi^*\xi\rangle\right.
\nonumber \\
&&\left.+d_3\langle\chi_a^*\chi_a\rangle
+\lambda_{\chi_sH}\langle H^\dagger H\rangle\right]
\langle\chi_s^*\rangle+a_3\langle\phi_s^\dagger\phi_a^\prime\rangle
+\left[c_{13}\langle\phi_a^\dagger\phi_s\rangle
+c_{22}\langle\phi_a^{\prime\dagger}\phi_s^\prime\rangle\right]
\langle\chi_a^*\rangle
\nonumber \\
&&+\left[c_{14}\langle\phi_s^\dagger\phi_a\rangle
+c_{23}^*\langle\phi_s^{\prime\dagger}\phi_a^\prime\rangle\right]
\langle\xi^*\rangle
+\left[a_5\langle\xi\rangle+c_{16}^*\langle\phi_a^{\prime\dagger}\phi_a\rangle
+c_{20}\langle\phi_s^{\prime\dagger}\phi_s\rangle
+d_4\langle\xi^*\xi^*\rangle\right]\langle\chi_a\rangle
\nonumber \\
&&+\left[c_{17}\langle\phi_s^{\prime\dagger}\phi_a\rangle
+d_6^*\langle\chi_a^*\chi_a^*\rangle\right]\langle\xi\rangle
+\left[3a_8\langle\chi_s\rangle
+2c_{19}\langle\phi_a^{\prime\dagger}\phi_s\rangle
+2d_5\langle\xi^*\chi_a^*\rangle\right]\langle\chi_s\rangle=0,
\label{eq:chis}
\\
&&\left[m_\xi^2+2\lambda_\xi\langle\xi^*\xi\rangle
+c_1\langle\phi_a^\dagger\phi_a\rangle+c_2\langle\phi_s^\dagger\phi_s\rangle
+c_3\langle\phi_a^{\prime\dagger}\phi_a^\prime\rangle
+c_4\langle\phi_s^{\prime\dagger}\phi_s^\prime\rangle
+d_1\langle\chi_a^*\chi_a\rangle+d_2\langle\chi_s^*\chi_s\rangle\right.
\nonumber \\
&&\left.+\lambda_{\xi H}\langle H^\dagger H\rangle\right]\langle\xi^*\rangle
+a_1\langle\phi_a^\dagger\phi_a^\prime\rangle
+a_4\langle\phi_s^\dagger\phi_s^\prime\rangle
+\left[c_{14}^*\langle\phi_a^\dagger\phi_s\rangle
+c_{23}\langle\phi_a^{\prime\dagger}\phi_s^\prime\rangle
+d_5^*\langle\chi_a\chi_s^*\rangle\right]\langle\chi_s^*\rangle
\nonumber \\
&&+\left[c_{15}^*\langle\phi_s^\dagger\phi_a\rangle
+c_{21}^*\langle\phi_s^{\prime\dagger}\phi_a^\prime\rangle\right]
\langle\chi_a^*\rangle
+\left[c_{17}\langle\phi_s^{\prime\dagger}\phi_a\rangle
+d_6^*\langle\chi_a^*\chi_a^*\rangle+a_5\langle\chi_a\rangle\right]
\langle\chi_s\rangle
+c_{18}^*\langle\phi_a^{\prime\dagger}\phi_s\rangle\langle\chi_a\rangle
\nonumber \\
&&+\left[3a_6\langle\xi\rangle
+2c_{25}^*\langle\phi_a^{\prime\dagger}\phi_a\rangle
+2c_{26}^*\langle\phi_s^{\prime\dagger}\phi_s\rangle
+2d_4^*\langle\chi_a^*\chi_s^*\rangle\right]\langle\xi\rangle=0,
\label{eq:xi}
\\
&&\left[m_{\phi_a}^2+2\lambda_{\phi_a}\langle\phi_a^\dagger\phi_a\rangle
+b_1\langle\phi_s^\dagger\phi_s\rangle
+b_2\langle\phi_a^{\prime\dagger}\phi_a^\prime\rangle
+b_3\langle\phi_s^{\prime\dagger}\phi_s^\prime\rangle
+c_1\langle\xi^*\xi\rangle+c_5\langle\chi_a^*\chi_a\rangle
+c_9\langle\chi_s^*\chi_s\rangle\right.
\nonumber \\
&&\left.+\lambda_{\phi_a H}\langle H^\dagger H\rangle\right]
\langle\phi_a^\dagger\rangle
+\left[a_1^*\langle\xi^*\rangle
+b_{10}\langle\phi_s^\dagger\phi_s^\prime\rangle
+b_{14}\langle\phi_a^\dagger\phi_a^\prime\rangle
+c_{16}^*\langle\chi_a\chi_s\rangle+c_{25}^*\langle\xi\xi\rangle\right]
\langle\phi_a^{\prime\dagger}\rangle
\nonumber \\
&&+\left[b_8\langle\phi_a^\dagger\phi_s\rangle
+b_9\langle\phi_a^{\prime\dagger}\phi_s^\prime\rangle
+c_{13}^*\langle\chi_s^*\chi_a\rangle
+c_{14}\langle\xi^*\chi_s\rangle+c_{15}^*\langle\chi_a^*\xi\rangle
\right]\langle\phi_s^\dagger\rangle
+\left[a_2^*\langle\chi_a^*\rangle
+b_7\langle\phi_a^\dagger\phi_s^\prime\rangle\right.
\nonumber \\
&&\left.+c_{17}\langle\chi_s\xi\rangle+c_{24}\langle\chi_a\chi_a\rangle
\right]\langle\phi_s^{\prime\dagger}\rangle=0,
\label{eq:phia}
\\
&&\left[m_{\phi_s}^2+2\lambda_{\phi_s}\langle\phi_s^\dagger\phi_s\rangle
+b_1\langle\phi_a^\dagger\phi_a\rangle
+b_4\langle\phi_a^{\prime\dagger}\phi_a^\prime\rangle
+b_5\langle\phi_s^{\prime\dagger}\phi_s^\prime\rangle
+c_2\langle\xi^*\xi\rangle+c_6\langle\chi_a^*\chi_a\rangle
+c_{10}\langle\chi_s^*\chi_s\rangle\right.
\nonumber \\
&&\left.+\lambda_{\phi_s H}\langle H^\dagger H\rangle\right]
\langle\phi_s^\dagger\rangle
+\left[a_3^*\langle\chi_s^*\rangle
+b_{11}\langle\phi_s^\dagger\phi_a^\prime\rangle
+c_{18}^*\langle\chi_a\xi\rangle+c_{19}\langle\chi_s\chi_s\rangle
\right]\langle\phi_a^{\prime\dagger}\rangle
+\left[a_4^*\langle\xi^*\rangle\right.
\nonumber \\
&&\left.+b_{10}^*\langle\phi_a^\dagger\phi_a^\prime\rangle
+b_{12}\langle\phi_s^\dagger\phi_s^\prime\rangle
+c_{20}\langle\chi_a\chi_s\rangle+c_{26}^*\langle\xi\xi\rangle
\right]\langle\phi_s^{\prime\dagger}\rangle
+\left[b_8\langle\phi_s^\dagger\phi_a\rangle
+b_9^*\langle\phi_s^{\prime\dagger}\phi_a^\prime\rangle
+c_{13}\langle\chi_a^*\chi_s\rangle\right.
\nonumber \\
&&\left.+c_{14}^*\langle\chi_s^*\xi\rangle
+c_{15}\langle\xi^*\chi_a\right]\langle\phi_a^\dagger\rangle=0,
\label{eq:phis}
\\
&&\left[m_{\phi_a^\prime}^2
+2\lambda_{\phi_a^\prime}\langle\phi_a^{\prime\dagger}\phi_a^\prime\rangle
+b_2\langle\phi_a^\dagger\phi_a\rangle+b_4\langle\phi_s^\dagger\phi_s\rangle
+b_6\langle\phi_s^{\prime\dagger}\phi_s^\prime\rangle
+c_3\langle\xi^*\xi\rangle+c_7\langle\chi_a^*\chi_a\rangle
+c_{11}\langle\chi_s^*\chi_s\rangle\right.
\nonumber \\
&&\left.+\lambda_{\phi_a^\prime H}\langle H^\dagger H\rangle\right]
\langle\phi_a^{\prime\dagger}\rangle+\left[a_1\langle\xi\rangle
+b_{10}^*\langle\phi_s^{\prime\dagger}\phi_s\rangle
+b_{14}\langle\phi_a^{\prime\dagger}\phi_a\rangle
+c_{16}\langle\chi_a^*\chi_s^*\rangle+c_{25}\langle\xi^*\xi^*\rangle
\right]\langle\phi_a^\dagger\rangle
\nonumber \\
&&+\left[a_3\langle\chi_s\rangle
+b_{11}\langle\phi_a^{\prime\dagger}\phi_s\rangle
+c_{18}\langle\chi_a^*\xi^*\rangle+c_{19}^*\langle\chi_s^*\chi_s^*\rangle
\right]\langle\phi_s^\dagger\rangle
+\left[b_{13}\langle\phi_a^{\prime\dagger}\phi_s^\prime\rangle
+b_9^*\langle\phi_a^\dagger\phi_s\rangle
+c_{21}^*\langle\xi\chi_a^*\rangle\right.
\nonumber \\
&&\left.+c_{22}^*\langle\chi_a\chi_s^*\rangle
+c_{23}^*\langle\chi_s\xi^*\rangle\right]\langle\phi_s^{\prime\dagger}\rangle=0,
\label{eq:phiap}
\\
&&\left[m_{\phi_s^\prime}^2
+2\lambda_{\phi_s^\prime}\langle\phi_s^{\prime\dagger}\phi_s^\prime\rangle
+b_3\langle\phi_a^\dagger\phi_a\rangle+b_5\langle\phi_s^\dagger\phi_s\rangle
+b_6\langle\phi_a^{\prime\dagger}\phi_a^\prime\rangle
+c_4\langle\xi^*\xi\rangle+c_8\langle\chi_a^*\chi_a\rangle
+c_{12}\langle\chi_s^*\chi_s\rangle\right.
\nonumber \\
&&\left.+\lambda_{\phi_s^\prime H}\langle H^\dagger H\rangle
\right]\langle\phi_s^{\prime\dagger}\rangle
+\left[a_2\langle\chi_a\rangle
+b_7\langle\phi_s^{\prime\dagger}\phi_a\rangle
+c_{17}^*\langle\chi_s^*\xi^*\rangle+c_{24}^*\langle\chi_a^*\chi_a^*\rangle
\right]\langle\phi_a^\dagger\rangle
+\left[a_4\langle\xi\rangle
+b_{10}\langle\phi_a^{\prime\dagger}\phi_a\rangle\right.
\nonumber \\
&&\left.+b_{12}\langle\phi_s^{\prime\dagger}\phi_s\rangle
+c_{20}^*\langle\chi_a^*\chi_s^*\rangle+c_{26}\langle\xi^*\xi^*\rangle
\right]\langle\phi_s^\dagger\rangle
+\left[b_9\langle\phi_s^\dagger\phi_a\rangle
+b_{13}\langle\phi_s^{\prime\dagger}\phi_a^\prime\rangle
+c_{21}\langle\xi^*\chi_a\rangle+c_{22}\langle\chi_a^*\chi_s\rangle\right.
\nonumber \\
&&\left.+c_{23}\langle\chi_s^*\xi\rangle\right]
\langle\phi_a^{\prime\dagger}\rangle=0.
\label{eq:phisp}
\end{eqnarray}

In section 6, in order to obtain correct phenomenology in the
neutrino sector,
the order magnitude of the vevs of different scalar fields of our model
have been estimated. Since these vevs are determined by the
Eqs. (\ref{eq:h}) $-$ (\ref{eq:phisp}), one has to adjust the unknown
parameters of these relations in such a way that the above mentioned order of
magnitude
for these vevs can be obtained. In order to fix these parameters, we
first assume that the mass-square parameters in Eq. (\ref{eq:pot})
should be around the square of the vevs of the corresponding fields. This
assumption is based on the fact that after spontaneous symmetry breaking,
a scalar field acquires mass around the scale at which the symmetry is
broken. As a result of this, we take the scales of the mass-square parameters
as:
\begin{eqnarray}
&&m_H^2\sim(100~{\rm GeV})^2,\quad m_{\chi_a}^2,m_{\chi_s}^2\sim(1~{\rm TeV})^2,
\quad m_{\phi_a}^2,m_{\phi_s}^2,m_{\phi_a^\prime}^2,m_{\phi_s^\prime}^2
\sim(10^{12}~{\rm GeV})^2,
\nonumber \\
&&m_\xi^2\sim(10^{17}~{\rm GeV})^2.
\end{eqnarray}
Now, in order to achieve the desired magnitude of the vevs for the scalar
fields, we can estimate the unknown parameters of
Eqs. (\ref{eq:h}) $-$ (\ref{eq:phisp}). These are given below.
\begin{eqnarray}
&&\lambda_H,\lambda_{\chi_a},\lambda_{\chi_s},\lambda_\xi,\lambda_{\phi_a},
\lambda_{\phi_s},\lambda_{\phi_a^\prime},\lambda_{\phi_s^\prime}\sim 1,
\nonumber \\
&& \lambda_{\phi_a H},\lambda_{\phi_s H},\lambda_{\phi_a^\prime H},
\lambda_{\phi_s^\prime H}\sim 10^{-20},\quad
\lambda_{\chi_a H},\lambda_{\chi_s H}\sim 10^{-2},\quad
\lambda_{\xi H}\sim 10^{-30},
\nonumber \\
&& a_1,a_4\sim 10^7~{\rm GeV},\quad a_2,a_3\sim 10^{-15}~{\rm GeV},\quad
a_5\sim 10^{-11}~{\rm GeV},\quad a_6\sim 10^{17}~{\rm GeV},\quad
\nonumber \\
&& a_7,a_8\sim 10^3~{\rm GeV},\quad b_1,\cdots,b_{14}\sim 1,
\nonumber \\
&& c_1,c_2,c_3,c_4,c_{25},c_{26}\sim 10^{-10},\quad
c_5,c_6,c_7,c_8,c_9,c_{10},c_{11},c_{12},c_{13},c_{16},c_{19},c_{20},c_{22},
c_{24}\sim 10^{-18},
\nonumber \\
&&c_{14},c_{15},c_{17},c_{18},c_{21},c_{23}\sim 10^{-32},\quad
d_1,d_2,d_4\sim 10^{-28},\quad d_3\sim 1,\quad d_5,d_6\sim 10^{-14}.
\end{eqnarray}
From the above mentioned values for the unknown parameters, we notice that some
couplings need to be suppressed in order to achieve the desired hierarchy in
the vevs of the scalar fields of our model.

Some parameters in the potential of Eq. (\ref{eq:pot}) can be complex.
As a result of this, after solving Eqs. (\ref{eq:h}) $-$ (\ref{eq:phisp}),
except for the Higgs field, rest of the fields can acquire complex vevs.
We will explain about the vacuum alignment for $\phi_a$ and $\phi_s$
shortly below. Assuming this vacuum alignment, using the above mentioned
complex vevs in our model of previous section,
Dirac and Majorana mass matrices for neutrinos are generated with
complex elements. Hence, the following parameters are complex:
$a$, $e$, $\epsilon_i$, $M_{\rm atm}$ and $M_{\rm sol}$. With these complex
parameters, we can explain the neutrino
masses and mixing angles and it is described in section 4 that the phases
of these
parameters can be fixed in order to get real values for the neutrino
masses. On the other hand, to explain the neutrino oscillation data,
it is sufficient to make $\epsilon_i$ to
be complex and rest of the above mentioned parameters can be chosen to be real.
To achieve
this particular case, we choose the following parameters of Eq. (\ref{eq:pot})
to be complex: $a_1,a_4,a_5,a_6,c_{14},c_{15},c_{17},c_{18},c_{21},c_{23},
c_{25},c_{26},d_4,d_5,d_6$. The phases of these parameters can be adjusted
in such a way that only $\langle\xi\rangle$ can be complex and rest of the
scalar fields have real vevs. As a result of this, only $\epsilon_i$
become complex and rest of the parameters of Dirac and Majorana
mass matrices can be real. It is explained before that the scalar fields
$\phi_a$ and $\phi_s$ need to acquire vevs in some particular directions,
in order to obtain TBM pattern in the neutrino sector. The vev for
these fields are determined by solving Eqs. (\ref{eq:phia}) $-$
(\ref{eq:phisp}). The unknown parameters in these equations should be
fine-tuned in such a way that the vevs for $\phi_a$ and $\phi_s$ can acquire the
desired directions. One can notice that, in order to do this fine-tuning,
enough number of parameters exist
in Eqs. (\ref{eq:phia}) $-$ (\ref{eq:phisp}). Hence, in our
scenario, the desired vacuum alignment for $\phi_a$ and $\phi_s$ can be
achieved. The scalar potential given in Eq. (\ref{eq:pot}) is at tree level.
This potential can get corrections at loop level. Since
loop effects give small corrections to the tree level potential and
due to large number of parameters in our model, we can still fine-tune these
parameters in order to get the necessary vacuum alignment for the above
scalar fields.

\end{document}